%% file: main.tex
\documentclass[acmsmall]{acmart}

\AtBeginDocument{%
  }

\usepackage{enumitem}
\usepackage{lipsum}
\usepackage{array}
\usepackage{wrapfig}
\usepackage{subcaption}
\usepackage{pgfplots}
\usepgfplotslibrary{groupplots}
\pgfplotsset{compat=1.17}
\usepackage{amsmath,amsfonts}
\usepackage[noend]{algpseudocode}
\usepackage{graphicx}
\usepackage{textcomp}
\usepackage{float}
\usepackage{listings}
\usepackage{xspace}
\usepackage{multirow}
\usepackage{amsthm}

\usepackage{balance}
\usepackage{algorithm}
\usepackage{algpseudocode}
\usepackage{colortbl}
\usepackage{subcaption}

\usepackage[skins]{tcolorbox}
\usepackage{xcolor}
\usepackage{multicol}
\usepackage{soul}
\usepackage{framed}
\usepackage{booktabs}

\usepackage{MnSymbol}

\usepackage{booktabs}
\usepackage{multirow}
\usepackage{siunitx}
\newcommand*\colourcheck[1]{%
	\expandafter\newcommand\csname #1check\endcsname{\textcolor{#1}{\ding{52}}}%
}
\colourcheck{blue}
\colourcheck{green}
\colourcheck{red}

\definecolor{custom-blue}{rgb}{0,0,0}

\newtcolorbox{boxB}[2][]{%
  enhanced,colback=white,colframe=black,coltitle=black,
  sharp corners,
  toprule=1.0pt,
  rightrule=0.3pt,
  leftrule=0pt,
  bottomrule=0pt,
  fonttitle=\itshape\scshape\large,
  left=0pt,right=5pt,top=5pt,bottom=3pt,
  attach boxed title to top right={yshift=-0.3\baselineskip-0.4pt,xshift=-5mm},
  boxed title style={tile,size=minimal,left=0.2mm,right=0.5mm,
    colback=white,before upper=\strut},
  title=#2,#1
}

\newtcolorbox{promptbox}[1][]{ 
  colback=lightgray!10,  
  colframe=gray,      
  fonttitle=\bfseries,    
  fontupper=\tiny,
  title={\tiny #1},
  before=\definecolor{royalblue}{rgb}{0.2549, 0.4118, 0.8824} 
}
\newcommand{\tool}{\textsc{AdaptAgent}\xspace}

\newboolean{showcomments}
\setboolean{showcomments}{true}
\ifthenelse{\boolean{showcomments}}
 { \newcommand{\mynote}[2]{
      \fbox{\bfseries\sffamily\scriptsize#1}
        {\small$\blacktriangleright$\textsf{\emph{#2}}$\blacktriangleleft$}}}
        { \newcommand{\mynote}[2]{}}

\usepackage{tikz}

\newcommand{\circlednum}[2][red]{%
\tikz[baseline=(char.base)]{
    \node[shape=circle, fill=#1, text=white, inner sep=1pt] (char) {#2};}%
}

\newtheorem{Key Idea}{Key Idea}

\newcolumntype{L}[1]{>{\raggedright\arraybackslash}p{#1}}

\newcommand{\code}[1]{{\footnotesize\texttt{#1}}}
\usepackage{amsthm}
\definecolor{dkgreen}{rgb}{0,0.6,0}
\definecolor{gray}{rgb}{0.5,0.5,0.5}
\definecolor{lightgray}{rgb}{211, 211, 211}
\definecolor{mauve}{rgb}{0.58,0,0.82}

\definecolor{dkgreen}{rgb}{0,0.6,0}
\definecolor{custom-red}{rgb}{1,0,0}
\definecolor{my-blue}{rgb}{0,0,1}
\definecolor{gray}{rgb}{0.5,0.5,0.5}
\definecolor{mauve}{rgb}{0.58,0,0.82}

\definecolor{c1}{HTML}{f4cccc}
\definecolor{c2}{HTML}{f5cdcd}
\definecolor{c3}{HTML}{fffcfc}
\definecolor{c4}{HTML}{ffffff}
\definecolor{c5}{HTML}{ffffff}
\definecolor{c6}{HTML}{fffdfd}
\definecolor{c7}{HTML}{f5cfcf}
\definecolor{c8}{HTML}{fffbfb}
\definecolor{c9}{HTML}{ffffff}
\definecolor{c10}{HTML}{fffdfd}
\definecolor{c11}{HTML}{fefafa}
\definecolor{c12}{HTML}{fef7f7}
\definecolor{c13}{HTML}{ffffff}
\definecolor{c14}{HTML}{fffefe}
\definecolor{c15}{HTML}{ffffff}
\definecolor{c16}{HTML}{fefafa}
\definecolor{c17}{HTML}{fdf3f3}
\definecolor{c18}{HTML}{fffefe}
\definecolor{c19}{HTML}{fdf5f5}
\definecolor{c20}{HTML}{ffffff}

\setcopyright{cc}
\setcctype{by-nc-nd}
\acmDOI{10.1145/3832284}
\acmYear{2026}
\acmJournal{PACMSE}
\acmVolume{3}
\acmNumber{ISSTA}
\acmArticle{ISSTA193}
\acmMonth{10}
\acmSubmissionID{issta26main-p2324-p}
\received{2026-01-30}
\received[accepted]{2026-06-25}

\begin{document}







\title{AdaptAgent: A Multi-agent, Domain-Guided Reasoning Framework for Code Adaptation}








\author{Xiaokai Rong}
\orcid{0009-0000-8457-8528}
\affiliation{%
  \institution{University of Texas at Dallas}
  \city{Dallas}
  \country{USA}
}
\email{xiaokai.rong@utdallas.edu}

\author{Hridya Dhulipala}
\orcid{0009-0001-4474-2984}
\affiliation{%
  \institution{University of Texas at Dallas}
  \city{Dallas}
  \country{USA}
}
\email{Hridya.Dhulipala@utdallas.edu}

\author{Aashish Yadavally}
\orcid{0000-0001-8785-6319}
\affiliation{%
  \institution{University of Central Florida}
  \city{Orlando}
  \country{USA}
}
\email{aashish.yadavally@ucf.edu}

\author{Tien N. Nguyen}
\orcid{0009-0006-7962-6090}
\affiliation{%
  \institution{University of Texas at Dallas}
  \city{Dallas}
  \country{USA}
}
\email{tien.n.nguyen@utdallas.edu}

\input{abstract-issta}

\ccsdesc[500]{Computing methodologies~Neural networks}
\ccsdesc[500]{Software and its engineering}

\keywords{AI4SE, Agentic SE, Multi-agents, Code Adaptation}





\maketitle

\input{intro-issta}

\input{motiv-issta}

\input{overview-issta}

\input{technical-section/summarization-phase-issta}

\input{technical-section/planning-phase-issta}

\input{technical-section/final-adaptation-phase}

\input{research-questions/overview}

\input{research-questions/rq1}
\input{research-questions/rq2}
\input{research-questions/rq3}

\input{research-questions/rq4}

\input{research-questions/rq5}
\input{limitations}

\input{related}

\input{conclusion}


\section{Data Availability}

Data/code is available in {\color{black}{project's website~\cite{adaptation2026}}}.

\newpage

\balance

\bibliographystyle{ACM-Reference-Format}

\bibliography{references}

\input{appendix}
\end{document}

%% file: abstract-issta.tex
\begin{abstract}


    

Developers often need to adapt into their projects the code generated from LLMs or code snippets from online forums. However, integrating them into an existing repository remains challenging in a manual process. 
A successful integration typically requires more than copying code as a user must produce correct adapting changes at a designated location in the target repository. We formalize this as the {\bf code adaptation problem}: given a snippet, functional intent, a target repository, and an adaptation location, generate a patch that adapts the snippet into the repository. We present {\tool}, a {\em multi-agent, domain-guided reasoning framework for code adaptation}. Rather than relying on single-shot prompting, {\tool} decomposes adaptation into specialized agents that communicate via typed artifacts: an \emph{Intent Summarizer} extracts adaptation goals from Q$\&$A text; a \emph{Policy Agent} derives domain policies from six adaptation categories; a \emph{Domain Planner} generates a self-ordered plan; a \emph{Context Miner} distills sibling-method semantics from the target codebase; and a \emph{Code Adapter} realizes the plan as a minimal unified diff, iteratively refined using a compiler-based \emph{Verifier}. This division of labor enables robust, policy-aligned adaptations and supports adapting code snippets into a project. On a real-world dataset, {\tool} outperforms strong baselines in semantic correctness and produces patches that mirror developers' actual adaptation patterns. Our ablation study shows each agent’s necessity, especially planning for code‑hardening and exception‑handling, and intent for logic customization.

\end{abstract}

%% file: intro-issta.tex
\section{Introduction}
\label{sec:intro}

Software developers frequently turn to LLMs or forums for quick solutions, with millions of code snippets shared daily~\cite{zhang-icse19,sadowski-fse15,10.1109/ICSE.2019.00046,verdi-tse22,hong21dicos,fisher17stack,chaiyong2021toxic,zhong2025developerllmconversationsempiricalstudy}. Currently, adapting these snippets requires developers to manually analyze, refactor, and integrate the online code to ensure correctness and compatibility. This process is both time-consuming and error-prone, requiring a deep understanding of both the snippet and the target codebase. 
An automated tool that facilitates correct and efficient adaptation would be highly beneficial, ensuring that adapted code is not only seamlessly integrated but also semantically correct within the project. In this work, we formalize the above scenarios as the {\bf code adaptation problem}: {\em given a code snippet, (functional) adaptation intent, a target repository, and an adaptation location, generate a change patch that adapts the snippet into the repository}. The adaptation intent in a forum is described via the questions and realized in the answers.

Large Language Models (LLMs) have shown strong performance on a range of code understanding and generation tasks.
However, their effectiveness on code adaptation remains unknown.
In our study, we explored LLMs’ capability for code adaptation and found that directly prompting an LLM to adapt a code snippet into a target project yields unsatisfactory results.
Our preliminary experiment (Section~\ref{sec:eval}) with direct prompting to GPT-4 revealed two primary shortcomings. First, a Stack Overflow (SO) post often contains irrelevant information, including unnecessary explanations, alternative solutions, and unrelated discussions. This is challenging for GPT-4 to extract the core adaptation intent, leading to incorrect and incomplete modifications. Second, when adapting code, GPT-4 frequently omits crucial elements such as code hardening (e.g., missing conditions, final modifiers, resource cleanup) and exception handling (e.g., \code{try-catch} blocks, exception declarations, or error propagation changes). These omissions can result in fragile or incorrect code.

In this paper, we present {\tool}, a multi‑agent framework that automates the adaptation of code snippets into a target project. {\tool} decomposes the task across specialized agents that communicate via typed artifacts. We leverage {\em 1) LLM-based adaptation planning, 2) domain-specific knowledge, and 3) context-aware adaptation} to ensure precise and reliable integration:

1. {\bf Adaptation Intent Summarizer} is an LLM agent that 
summarizes the essential parts of the SO post, filtering out irrelevant details while retaining the {\em adaptation goal} from noisy texts.

2. {\bf Strategic Adaptation Planner} is an LLM agent in which instead of performing adaptation in a single step, it employs autonomous planning for an LLM to generate a self-ordered adaptation plan
before {\em executing code modifications}. 
The plan includes identifying necessary modifications based on the post and the target codebase, and structuring a step-by-step adaptation and validation checks to detect potential issues before finalizing the adaptation -- all autonomously planned by the agent. To guide the planner agent, {\tool} uses a {\em policy agent} that derives the adaptation policy from domain knowledge with six adaptation categories.

3. {\bf Policy Agent} is an agent that performs domain-specific knowledge integration.
Prior research in Zhang {\em et al.}~\cite{10.1109/ICSE.2019.00046} has identified six major categories of code adaptation: {\em code hardening, resolving compilation errors, exception handling, logic customization, refactoring, and miscellaneous improvements}. The policy agent incorporates this domain knowledge as well as any project-specific policies into a checklist, guiding the LLM to systematically address these common adaptation challenges, ensuring that the generated code aligns with best practices. 

4. {\bf Context Miner} is an {\em external tool} to extract the adaptation context.
{\color{custom-blue}{To help {\tool} understand the target codebase for adaptation}}, we provide
the target method/class context.


5. {\bf Code Adapter} is a {\em unified‑diff generation agent} realizing the plan as a minimal diff. With~planning, LLM {\em autonomously plans the potentially conflicting atomic adaptation changes, achieving~a self-ordered adaptation plan}. We do not need to specify the order of concrete  adaptation~changes.

6. {\bf Verifier} is {\em an external compiler} to provide the feedback to the adapter to {\color{custom-blue}{correct any syntactic errors}} in the adapted~code.


Our empirical evaluation on a real-world code adaptation dataset shows that {\tool} achieves 63.1\% semantic correctness, which is {\color{custom-blue}{5.5\%-53.6\% more correct adaptation cases than the baselines}}. We also demonstrated that the code adapted by {\tool} {\color{custom-blue}{covers similar adaptation aspects in proportions similar to the code adapted by actual developers}} in their projects. The adaptation correctness is also consistent across adaptation complexity. Finally, we showed that all four components (intent, policy guidance, planning, and context) contributed positively to {\tool}'s performance. In brief, in this paper, we make the following contributions:

1. {\bf {\tool}}: A {\bf multi‑agent, domain‑guided framework} for context‑aware code adaptation with explicit agent contracts {\color{custom-blue}{via a linear pipeline coordination with typed artifacts}};

2. {\bf Autonomous code adaptation} across categories without predefined order of atomic changes;

3. {\bf Extensive empirical evaluation} to show the effectiveness of {\tool} in code adaptation over the baselines and the contributions of each component to the overall performance.

%% file: motiv-issta.tex
\section{Motivation}
\label{sec:motiv}




Let us present a real-world example to motivate our approach. A developer posted a question on StackOverflow (SO) with the post ID \#29764243~\cite{so-post29764243}, stating: {\em "I am trying to draw and erase on image and then save it to sd card. 
I got success in drawing and saving it to sd card, but when I checked 
the save image in sd card I got a black background image not the image that I drew. Please suggest how to apply an image as a canvas background and erase the drawn text without
losing the background image. I have seen various examples on stack overflow but my bad nothing is working for me, pls help."} (S)he also provided the relevant code to clarify the context of the question. 

\begin{wrapfigure}{r}{3.2in}
	\centering
    \definecolor{gray}{rgb}{0.5,0.5,0.5}
    \definecolor{mauve}{RGB}{127,0,145}
    \definecolor{lightgray}{gray}{0.97}
    \lstset{
        mathescape=true,
        backgroundcolor=\color{lightgray},
    	numbersep=-4pt,
    	keywordstyle=\color{mauve},
        basicstyle=\scriptsize\ttfamily,
        numberstyle=\scriptsize\ttfamily\color{gray},
        numbers=left,
        keywordstyle= \color{blue!70},
        commentstyle= \color{red!50!green!50!blue!50},
        rulesepcolor= \color{red!20!green!20!blue!20},
		xleftmargin=1.5em,xrightmargin=1em, aboveskip=1em,
		framexleftmargin=1.5em,
        numbersep= 5pt,
        language=Python,
        emphstyle=\bfseries,
        moredelim=**[is][\color{red}]{@}{@},
        escapeinside= {(*@}{@*)},
        breaklines=true,
        breakautoindent=false,
    }
\begin{lstlisting}[]
public class DrawingView extends View {
  ...
  private void touch_move(float x, float y) {
    float dx = Math.abs(x - mX);
    float dy = Math.abs(y - mY);
    if (dx >= TOUCH_TOLERANCE || dy >= TOUCH_TOLERANCE) {
        mPath.quadTo(mX, mY, (x + mX) / 2, (y + mY) / 2);
        mX = x;
        mY = y;

        circlePath.reset();
        circlePath.addCircle(mX, mY, 30, Path.Direction.CW);
        mCanvas.drawPath(mPath, mPaint);
    }
  } ...
\end{lstlisting}
\vspace{-16pt}
\caption{An Answer to the Question in SO Post \#29764243}
\label{motiv:example}
\end{wrapfigure}

Another developer replied with a solution, including the code in Fig.~\ref{motiv:example}. Later, the \code{Touch\_Move} method from the solution in SO was adapted into a GitHub project named \code{JustWeTools}~\cite{justwetools-github}.


In \code{JustWeTools} (Fig.~\ref{motiv:adapt}), the developers made a few modifications to adapt the original code to their specific needs. 
For instance, the code on lines 11–13 of the original answer in Fig.~\ref{motiv:example} was intended for demonstration purposes in the SO post and was therefore removed by the \code{JustWeTools} developers. Notably, they added the lines 10–14 (Fig.~\ref{motiv:adapt}) to implement a feature that enables {\em "applying an image as a canvas background and erasing or drawing text without losing the background image."} However, they {\em overlooked exception handling} for \code{drawPath} on lines 11 and 13 (Fig.~\ref{motiv:adapt}), which {\em could cause exceptions} if the image file path is~invalid.


As shown, code adaptation to an existing project is error-prone, necessitating an automated solution, yet no tool exists for seamless adaptation. Advancements in~LLMs present a new direction. To assess that, we conducted a preliminary experiment using GPT-4 with a simple prompt, providing (1) the question showing the (functional) adaptation intent of the code, and (2) the \code{Touch\_Move} method from the accepted answer. However, GPT-4 failed to adapt the code. With adjusted access control and a defined constant, the resulting code compiled but was functionally incorrect, {\em retaining irrelevant operations} (e.g., \code{addCircle} in lines 11-13, Fig.~\ref{motiv:example}) while {\em omitting functions} like canvas background change, text erasure without image loss (lines 10–14, Fig.~\ref{motiv:adapt}), and missing {\em exception handling}.

\input{motiv-example}

\section{Key Ideas}

The naive prompt clearly lacked sufficient guidance for the LLM in code adaptation. First, from the example, we can identify 
that the content of the posts and discussions surrounding the questions and answers can provide valuable insights into the functional adaptation intent. However, the entire post may include irrelevant information that could mislead the LLM in capturing the adaptation intent. For instance, while most of the question’s content is relevant, the last sentence is not. Thus, we hypothesize that a summarized version of the question and correct answers would be more concise and effective in conveying the intent for adaptation, which is expressed in Q\&A texts. For LLM-generated code, adaptation intent can be expressed in users' textual descriptions. 



\begin{Key Idea}[Adaptation Intent]
(Functional) adaptation intent in adapting code snippets can be effectively conveyed through a summary of the question and its answers. 
\end{Key Idea}



Second, to ensure a complete adaptation process, we propose integrating domain-specific knowledge. Specifically, Zhang {\em et al.}~\cite{10.1109/ICSE.2019.00046} conducted a large-scale study on the adaptations and variations of SO snippets by qualitatively analyzing 400 SO examples and the GitHub counterparts that adapted them. They developed a taxonomy of six categories of code adaptation: {\em code hardening, resolving compilation errors, exception handling, logic customization, refactoring, and miscellaneous}. 
We will incorporate the guidance in those categories into the prompts provided to the LLMs.


\begin{Key Idea}[Domain-Specific Guidance]
Guidance for LLMs can be developed effectively based on domain-specific knowledge collected from practice, e.g., online code-snippet adaptation knowledge.
\end{Key Idea}



Third, the target code's context is a critical factor in guiding adaptation. For example, the~\code{Touch}\code{\_Move} method (Fig.~\ref{motiv:adapt}) was integrated into the \code{DrawPath} class of the \code{JustWeTools} project. Providing details on the adaptation location, along with relevant semantic context, e.g., the other methods in the same class {\em (sibling methods)}, can be helpful. 
In this case, sibling methods like {\code{Touch\_Down}, \code{Touch\_Up}, and \code{onDraw} offer related functionality and share common APIs, providing valuable adaptation guidance.

\begin{Key Idea}[Semantic Context]
{\color{custom-blue}{The methods in the target class, into which the given code snippet is integrated, offer semantic context that helps guide the LLM in performing the adaptation.}}
\end{Key Idea}

Finally, adapting code snippets to an existing project 
involves understanding the project's context, dependencies, and developers' intentions.
Single-shot prompting may overlook crucial details or produce inconsistent results. In contrast, {\em multi-agent architecture} enables {\em cooperating specialized agents and/or external tools} with division of work in which LLM planning offers an {\em autonomous} approach, enabling it to break down the adaptation task into smaller reasoning/action steps. 



\begin{Key Idea}[Why Multi-agent and LLM Planning?]
Multi-agent architecture enables cooperating agents in which LLM planning leverages a more strategic process that integrates context, relevant guidance, and intermediate validations,
deriving a sequence of non-conflicting atomic changes that collectively satisfy all adaptation requirements. By reasoning over dependencies among atomic adaptation changes, the agents determine the right order of atomic changes without relying on a~pre-defined workflow, ensuring the final patch is correct, complete, and consistent with the target codebase. 

\end{Key Idea}


Fig.~\ref{motiv:our-adapt} illustrates the adaptation suggested by {\tool}, which leverages its multi-agent framework equipped with adaptation intent, domain-specific guidance, and semantic context. As shown, the code at line 11 in Fig.~\ref{motiv:our-adapt} features a ternary operator expression, equivalent to the conditional statement in the adapted code at lines 10–14 of Fig.~\ref{motiv:adapt}. Exception handling was added to gracefully manage drawing issues (lines 12–14, Fig.~\ref{motiv:our-adapt}). 



%% file: motiv-example.tex
\begin{figure}[t]
\begin{minipage}[t]{0.48\textwidth}
\centering
    \definecolor{gray}{rgb}{0.5,0.5,0.5}
    \definecolor{mauve}{RGB}{127,0,145}
    \definecolor{lightgray}{gray}{0.97}
    \lstset{
        mathescape=true,
        backgroundcolor=\color{lightgray},
    	numbersep=-4pt,
    	keywordstyle=\color{mauve},
        basicstyle=\scriptsize\ttfamily,
        numberstyle=\scriptsize\ttfamily\color{gray},
        numbers=left,
        keywordstyle= \color{blue!70},
        commentstyle= \color{red!50!green!50!blue!50},
        rulesepcolor= \color{red!20!green!20!blue!20},
		xleftmargin=1.5em,xrightmargin=1em, aboveskip=1em,
		framexleftmargin=1.5em,
        numbersep= 5pt,
        language=Python,
        emphstyle=\bfseries,
        moredelim=**[is][\color{red}]{@}{@},
        escapeinside= {(*@}{@*)},
        breaklines=true,
        breakautoindent=false,
    }
\begin{lstlisting}[]
public void Touch_Move(float x, float y) {
  super.Touch_Move(x, y);
  float dx = Math.abs(x - mX);
  float dy = Math.abs(y - mY);
  if (dx >= PaintView.TOUCH_TOLERANCE || dy >= PaintView.TOUCH_TOLERANCE) {
      mPath.quadTo(mX, mY, (x + mX) / 2, (y + mY) / 2);
      mX = x;
      mY = y;
      
      if(PaintView.IsPaint) {
         mCanvas.drawPath(mPath, mPaint);
      } else {
         mCanvas.drawPath(mPath, mEraserPaint);
      }
  }
}
\end{lstlisting}
\vspace{-16pt}
\caption{Human-adapted Code in \code{JustWeTools} Project}
\label{motiv:adapt}
\end{minipage}
\hfill
\begin{minipage}[t]{0.51\textwidth}
\centering
    \definecolor{gray}{rgb}{0.5,0.5,0.5}
    \definecolor{mauve}{RGB}{127,0,145}
    \definecolor{lightgray}{gray}{0.97}
   \lstset{
        mathescape=true,
        backgroundcolor=\color{lightgray},
    	numbersep=-4pt,
    	keywordstyle=\color{mauve},
        basicstyle=\scriptsize\ttfamily,
        numberstyle=\scriptsize\ttfamily\color{gray},
        numbers=left,
        keywordstyle= \color{blue!70},
        commentstyle= \color{red!50!green!50!blue!50},
        rulesepcolor= \color{red!20!green!20!blue!20},
		xleftmargin=1.5em,xrightmargin=1em, aboveskip=1em,
		framexleftmargin=1.5em,
        numbersep= 5pt,
        language=Python,
        emphstyle=\bfseries,
        moredelim=**[is][\color{red}]{@}{@},
        escapeinside= {(*@}{@*)},
        breaklines=true,
        breakautoindent=false,
    }
\begin{lstlisting}[]
public void Touch_Move(float x, float y) {
  super.Touch_Move(x, y);
  float dx = Math.abs(x - mX);
  float dy = Math.abs(y - mY);
  if (dx >= PaintView.TOUCH_TOLERANCE || dy >= PaintView.TOUCH_TOLERANCE) {
      mPath.quadTo(mX, mY, (x + mX) / 2, (y + mY)/2);
      mX = x;
      mY = y;

      try {
        mCanvas.drawPath(mPath, PaintView.IsPaint ? mPaint : mEraserPaint);
      } catch (Exception e) {        
        e.printStackTrace();
      }  
  }
}
\end{lstlisting}
\vspace{-16pt}
\caption{The Code Adapted by {\tool}}
\label{motiv:our-adapt}
\end{minipage}
\end{figure}

%% file: overview-issta.tex
\section{{\tool} Overview}
\label{sec:overview}

\begin{figure}[t]
  \centering
  \includegraphics[width=0.84\textwidth]{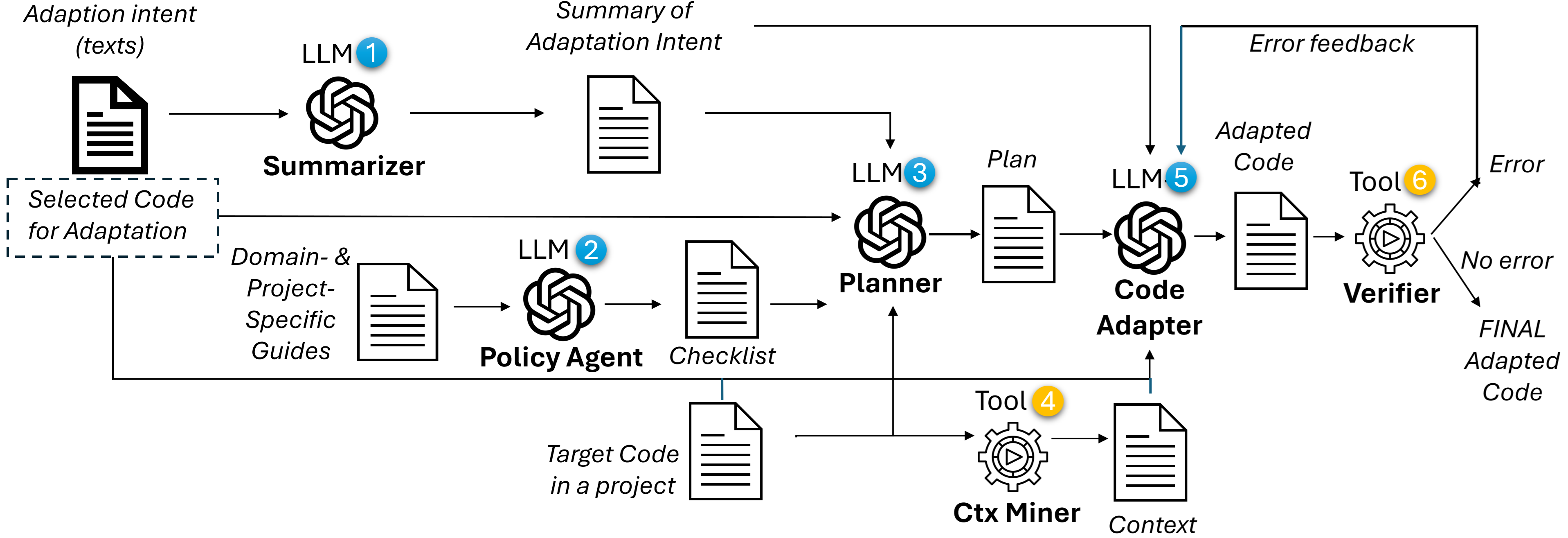} 
  \vspace{-9pt}
  \caption{{\tool}: A Multi-agent, Domain-Guided Reasoning Framework for Code Adaptation}
  \label{fig:overview}
\end{figure}

Fig.~\ref{fig:overview} illustrates {\tool}'s multi-agent architecture comprising of cooperating agents connected by typed artifacts. It takes as inputs the given code snippet, (functional) adaptation intent, the adaptation code location in the target codebase. For online code adaptation, the adaptation intent is expressed in the question and answer and the given code is the one in the accepted answer. 


First, the Summarizer (\circlednum[blue]{1}) captures the intent from the post. It takes as inputs the question, the accepted code snippet selected for adaptation, and the associated texts, among multiple potential answers. The Summarizer creates the summary, concisely capturing the {\em intent of the adaptation}. 


Second, the Policy agent (\circlednum[blue]{2}) yields a checklist from the domain-specific guide and any project-specific policies (e.g., error-handling style). For online code adaptation,
the common adaptation types include 1) code hardening (i.e., adding a conditional or a final modifier, cleaning up un-managed resources, etc.), 2) resolving compilation errors, 3) exception handling (i.e., inserting/deleting a \code{try-catch} block or a thrown exception in a method header, changing statements in \code{catch} or \code{finally} blocks, updating exception types), 4) logic customization, 5) refactoring (i.e., renaming, replacing magic numbers, inlining a field, etc.), and 6) others~\cite{10.1109/ICSE.2019.00046}. This agent builds the checklist that instructs the Planner to look for the opportunities to support those adaptation~categories.

Third, {\tool} leverages the Planner agent (\circlednum[blue]{3}) to develop an adaptation plan using the checklist produced by the Policy agent. In total, the Planner takes as inputs 1) the summary of the adaptation intent produced by the Summarizer agent, 2) the adaption checklist produced by the Policy agent, 3) the target code in the project (the adapting location), and 4) the selected code for adaptation. The output of the Planner is the adaptation plan with descriptions of change~actions.


Fourth, Context Miner component (\circlednum[orange]{4}) distills the important contextual information surrounding the adaption location from the target project for adaptation. In the current design, for simplification, we collect only the API method calls in the sibling methods as the semantic~context. However, our general framework can accommodate richer structural/semantic signals (types, call graphs, PDGs) that are often critical for correct integration. We left these richer signals for future work as {\tool} achieved high accuracy in code adaptation in our evaluation (Section~\ref{sec:eval}). 

Fifth, the Code Adapter agent (\circlednum[blue]{5}) takes as inputs 1) the adaptation plan, 2) the summary of the intent, 3) the target code and adaptation location in the target project, and 4) the semantic context. This agent outputs the adaptation patch. 
The Adapter LLM {\em autonomously determines the sequence of non-conflicting, atomic changes across relevant categories}, without explicitly specifying their order (see Fig.~\ref{motiv:our-adapt}). {\em This is the benefit of the use of multi-agents}.
Moreover, we use the Verifier (\circlednum[orange]{6}), which is an external compiler to provide the feedback to the Adapter to correct the adapted code. This feedback is repeated until no more syntactic errors or a limit number of attempts is reached.
{\color{custom-blue}{The agents use a linear pipeline coordination via typed messages including intent summary, policy checklist, self‑ordered adaptation plan, semantic context, unified diff, etc. (see artifacts in Fig.~\ref{fig:overview})}}.

%% file: technical-section/summarization-phase-issta.tex
\section{LLM-Based Content Summarization}



{\color{custom-blue}{Leveraging LLMs' capability in summarization}}~\cite{luo2023chatgptfactualinconsistencyevaluator,tian2023chatgptultimateprogrammingassistant}, Summarizer (\circlednum[blue]{1}) creates the summary of adaptation intent from a post. We formulate this step of abstracting adaptation intent,~\(\mathcal{I}\):
$\mathcal{M} (\mathcal{I} | \mathcal{P}_\mathcal{Q}, \mathcal{P}_\mathcal{C}, \mathcal{P}_\mathcal{A})$.
Given a post \(\mathcal{P}\) in our problem statement, we denote the question text, including any code or natural language descriptions as \(\mathcal{P}_\mathcal{Q}\), the specific answer text which has online answering code snippet that has been chosen by a questioner to be adapted into his/her codebase (possibly with an explanatory text), as \(\mathcal{P}_\mathcal{C}\), and the associated tags as \(\mathcal{P}_\mathcal{A}\).
 

The tuple \( \mathcal{P} = \langle \mathcal{P}_\mathcal{Q}, \mathcal{P}_\mathcal{C}, \mathcal{P}_\mathcal{A}\rangle  \) is consumed by the Summarizer (\circlednum[blue]{1}), leveraging LLM's capability to convert a mixture of natural language (NL) and programming language (PL) into a concise, high-level NL summary \(\mathcal{I}\). Specifically, \(\mathcal{P}_\mathcal{Q}\) often contains questions, while \(\mathcal{P}_\mathcal{C}\) represents the online code solution that a user aims to adapt. By combining both into a single prompt, Summarizer can make a connection between the question \(\mathcal{P}_\mathcal{Q}\) and the answer code snippet \(\mathcal{P}_\mathcal{C}\), thus capturing the adaptation intent. The tags \(\mathcal{P}_\mathcal{A}\) provide other meta-information, e.g., language, version, library constraints,~etc. Notably, these tags, such as {\em "Java 1.8"}, {\em "Apache 2.0"}, or even {\em "Debugging"}, encode contextual details on the environment or constraints that might be critical for an effective adaptation.


\input{figures/technical-section-prompt-example}

To capture the adaptation intent in a post, we construct a specialized prompt used in the Summarizer as shown in top left (\(\bigotimes\)) of Fig.~\ref{fig:all-prompt}. In this prompt, we explicitly guide LLM to structurally break down the core component in a post, including a descriptive title of the question, the primary problem statement, and notable constraints or environment details, the solution strategy within the chosen snippet, and any edge cases that possibly could be mentioned in the post. \( \mathcal{P} = \langle \mathcal{P}_\mathcal{Q}, \mathcal{P}_\mathcal{C}, \mathcal{P}_\mathcal{A} \rangle  \) is embedded into the location marked with \code{<POST>} in the prompt.

Within the prompt, for the sections {\em "Accepted Answer's Core Idea"} and {\em "Detailed Code Snippet Explanation,"} we guide the model to identify the most relevant connections from \(\mathcal{P}_\mathcal{Q} \rightarrow \mathcal{P}_\mathcal{C} \). For example, as shown in the bottom-left subfigure of Fig.~\ref{fig:all-prompt} (\( \oast \)), the sample output of our prompt—after applying to the motivation example—shows that the LLM successfully extracts the intent behind \(\mathcal{P}_\mathcal{Q} \rightarrow \mathcal{P}_\mathcal{C} \). Specifically, it identifies the core of the question ({\em "enabling a user to draw and erase parts of an image without losing the background in an Android drawing application"}) and the key idea of the code snippet in the solution (See {\em "Accepted Answer's Core Idea"}). The {\em "Core Problem"} section highlights the issue in incorrect background setting. \(\mathcal{P}_\mathcal{T}\) specifies Android app as the~domain.

%% file: figures/technical-section-prompt-example.tex
\begin{figure*}
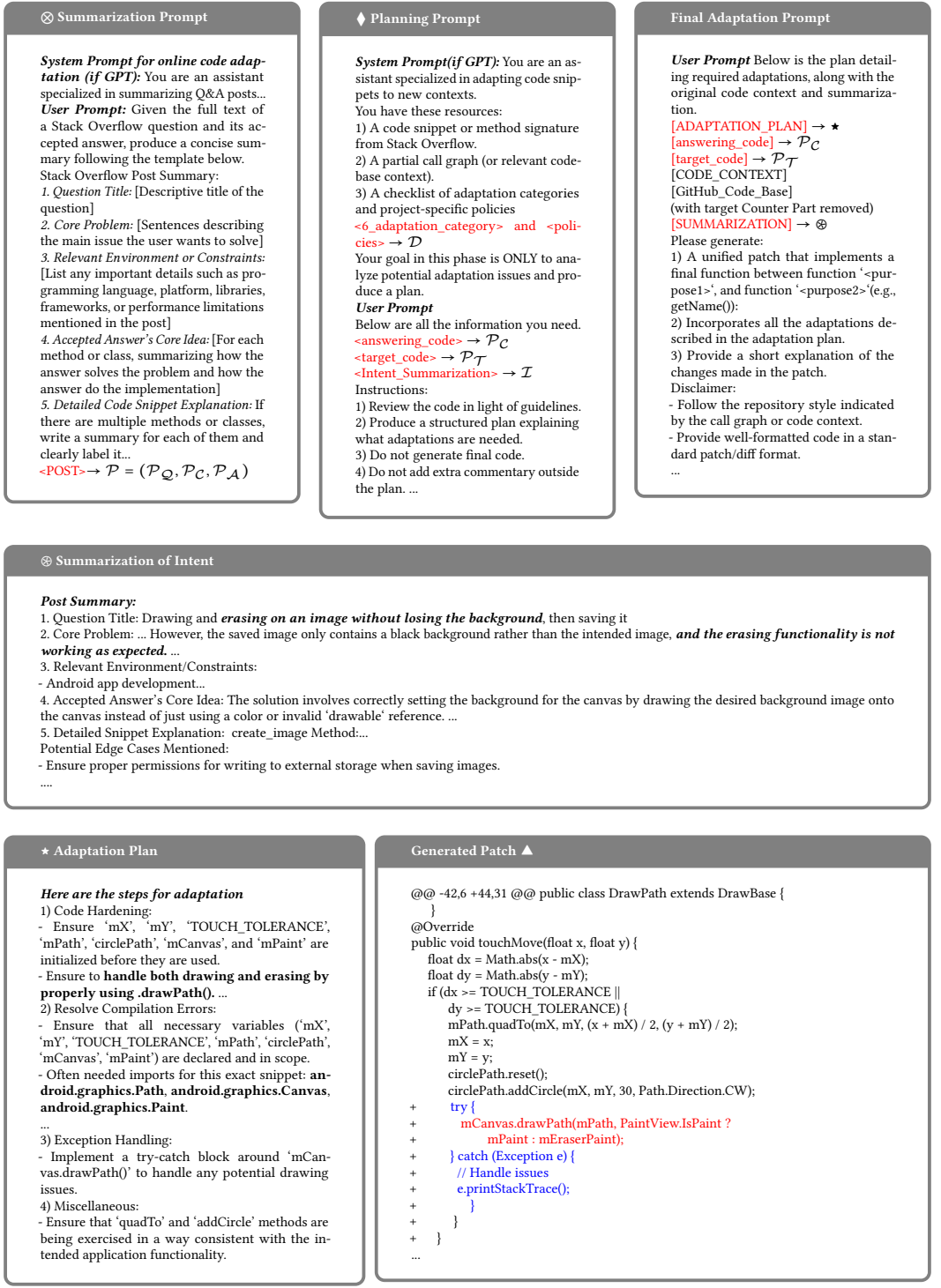

    \centering

    \begin{minipage}[t]{0.32\linewidth}\vspace{0pt}
        \begin{promptbox}[\(\bigotimes\) Summarization Prompt]
            {\bf\em System Prompt for online code adaptation (if GPT):}
            You are an assistant specialized in summarizing Q\&A posts...

            {\bf\em User Prompt:}
            Given the full text of a Stack Overflow question and its accepted answer, produce a concise summary following the template below.
            
            Stack Overflow Post Summary:
            
            {\em 1. Question Title:} [Descriptive title of the question]
            
            {\em 2. Core Problem:} [Sentences describing the main issue the user wants to solve]
            
            {\em 3. Relevant Environment or Constraints:} [List any important details such as programming language, platform, libraries, frameworks, or performance limitations mentioned in the post]
            
            {\em 4. Accepted Answer's Core Idea:} [For each method or class, summarizing how the answer solves the problem and how the answer do the implementation]
            
            {\em 5. Detailed Code Snippet Explanation:} If there are multiple methods or classes, write a summary for each of them and clearly label it... 
            
            \textcolor{red}{<POST>}\(\rightarrow \mathcal{P = (\mathcal{P}_\mathcal{Q}, \mathcal{P}_\mathcal{C},\mathcal{P}_A)}\)
            
        \end{promptbox}
    \end{minipage}
    \hfill
    \begin{minipage}[t]{0.32\linewidth}\vspace{0pt}
        \begin{promptbox}[\(\blacklozenge\) Planning Prompt ]

        {\bf\em System Prompt(if GPT):}
            You are an assistant specialized in adapting code snippets to new contexts. 
            
            You have these resources:
            
            1) A code snippet or method signature from Stack Overflow.
            
            2) A partial call graph (or relevant codebase context).
            
            3) A checklist of adaptation categories and project-specific policies
            
            \textcolor{red}{<6\_adaptation\_category> and <policies>} \(\rightarrow\mathcal{D}\)
            
            Your goal in this phase is ONLY to analyze potential adaptation issues and produce a plan. 
            
            {\bf\em User Prompt}
            
            Below are all the information you need.
            
            \textcolor{red}{<answering\_code>} \(\rightarrow\mathcal{P}_\mathcal{C}\)

            \textcolor{red}{<target\_code>} \(\rightarrow\mathcal{P}_\mathcal{T}\)
            
           \textcolor{red}{<Intent\_Summarization>} \(\rightarrow\mathcal{I}\)
            
            Instructions:
            
            1) Review the code in light of guidelines. 
            
            2) Produce a structured plan explaining what adaptations are needed. 
               
            3) Do not generate final code. 
            
            4) Do not add extra commentary outside the plan. ...
        \end{promptbox}
    \end{minipage}
    \hfill
    \begin{minipage}[t]{0.32\linewidth}\vspace{0pt}
        \begin{promptbox}[Final Adaptation Prompt]

            {\bf\em User Prompt}
            Below is the plan detailing required adaptations, along with the original code context and summarization.

            \textcolor{red}{[ADAPTATION\_PLAN]} \( \rightarrow \filledstar \)
            
            \textcolor{red}{[answering\_code]}    \( \rightarrow \mathcal{P}_\mathcal{C} \)

            \textcolor{red}{[target\_code]}    \( \rightarrow \mathcal{P}_\mathcal{T} \)
            
            [CODE\_CONTEXT]

            [GitHub\_Code\_Base] 
            
            (with target Counter Part removed)
            
            \textcolor{red}{[SUMMARIZATION]}  \( \rightarrow\oast \)
            
            Please generate:
            
            1) A unified patch that implements a final function between function `<purpose1>`, and function `<purpose2>`(e.g., getName()):
            
            2) Incorporates all the adaptations described in the adaptation plan.
            
            3) Provide a short explanation of the changes made in the patch.
            
            Disclaimer:
            
            - Follow the repository style indicated by the call graph or code context.
            
            - Provide well-formatted code in a standard patch/diff format.

            ...
        \end{promptbox}
    \end{minipage}

    \vspace{1em}  

    \begin{minipage}[t]{\linewidth}\vspace{0pt}
        \begin{promptbox}[ \( \oast \) Summarization of Intent]

            {\bf\em Post Summary:}
            
            1. Question Title: Drawing and {\bf\em erasing on an image without losing the background}, then saving it
            
            2. Core Problem: ...  However, the saved image only contains a black background rather than the intended image, {\bf\em and the erasing functionality is not working as expected.} ... 
            
            3. Relevant Environment/Constraints: 
            
               - Android app development...
            
            4. Accepted Answer's Core Idea: The solution involves correctly setting the background for the canvas by drawing the desired background image onto the canvas instead of just using a color or invalid `drawable` reference. ... 
            
            5. Detailed Snippet Explanation:
            \- create\_image Method:...
            
            Potential Edge Cases Mentioned:
            
            - Ensure proper permissions for writing to external storage when saving images.

            ....
        \end{promptbox}
    \end{minipage}
    \hfill
    \begin{minipage}[t]{0.39\linewidth}\vspace{10pt}
        \begin{promptbox}[\( \filledstar \) Adaptation Plan]

        {\bf\em Here are the steps for adaptation}
        
        1) Code Hardening:
        
           - Ensure `mX', `mY', `TOUCH\_TOLERANCE', `mPath', `circlePath', `mCanvas', and `mPaint' are initialized before they are used.
           
           - Ensure to {\bf handle both drawing and erasing by properly using .drawPath().} ... 

        2) Resolve Compilation Errors:
        
           - Ensure that all necessary variables (`mX', `mY', `TOUCH\_TOLERANCE', `mPath', `circlePath', `mCanvas', `mPaint') are declared and in scope.

           - Often needed imports for this exact snippet: \textbf{android.graphics.Path}, \textbf{android.graphics.Canvas}, \textbf{android.graphics.Paint}.
           
           ...           

        3) Exception Handling:
        
           - Implement a try-catch block around `mCanvas.drawPath()' to handle any potential drawing issues.

        4) Miscellaneous:
        
           - Ensure that `quadTo' and `addCircle' methods are being exercised in a way consistent with the intended application functionality.
            
        \end{promptbox}
    \end{minipage}
    \hfill
    \begin{minipage}[t]{0.6\linewidth}\vspace{10pt}
        \begin{promptbox}[Generated Patch \(\blacktriangle\)]
@@ -42,6 +44,31 @@ 
public class DrawPath extends DrawBase \{

\hspace{0.25cm}     \}
 
@Override

public void touchMove(float x, float y) \{

\hspace{0.25cm}float dx = Math.abs(x - mX);

\hspace{0.25cm}float dy = Math.abs(y - mY);

\hspace{0.25cm}if (dx >= TOUCH\_TOLERANCE ||

\hspace{0.5cm} dy >= TOUCH\_TOLERANCE) \{

\hspace{0.5cm}            mPath.quadTo(mX, mY, (x + mX) / 2, (y + mY) / 2);

\hspace{0.5cm}            mX = x;

\hspace{0.5cm}            mY = y;

\hspace{0.5cm}            circlePath.reset();

\hspace{0.5cm}            circlePath.addCircle(mX, mY, 30, Path.Direction.CW);

+\hspace{0.448cm}            \textcolor{blue}{try \{}

+ \hspace{0.55cm}                \textcolor{red}{mCanvas.drawPath(mPath, PaintView.IsPaint ?}

+      \hspace{0.95cm}                                      \textcolor{red}{mPaint : mEraserPaint);}

+\hspace{0.448cm}            \textcolor{blue}{\} catch (Exception e) \{}

+\hspace{0.55cm}             \textcolor{blue}{     // Handle issues}

+\hspace{0.55cm}            \textcolor{blue}{     e.printStackTrace();}

+\hspace{0.75cm}           \textcolor{blue}{\}}

+\hspace{0.5cm}        \}

+\hspace{0.25cm}    \}

...
        \end{promptbox}
    \end{minipage}
    \vspace{-6pt}
    \caption{Prompt Templates used in {\tool}}
    \label{fig:all-prompt}
\end{figure*}

%% file: technical-section/planning-phase-issta.tex
\section{Domain-Specific Knowledge-Guided Planning}
\label{sec:planning}


In the LLM planning, we provide the Planner agent with three core inputs to generate a structured adaptation plan. First, we integrate relevant domain-specific checklist \(\mathcal{D}\) derived from the Policy agent on adaptation categories, allowing the LLM to systematically determine the necessary adaptation modifications to the given target code snippet. Second, we incorporate the adaptation intent \(\mathcal{I}\) from the summarizer agent, ensuring modifications align with the overarching goal rather than isolated code changes. Third, we supply the target code itself. Finally, the LLM Planner produces a step-by-step adaptation plan \(\mathcal{Y}\), outlining actionable items \(\mathcal{A}\) and reasoning steps \(\mathcal{R}\).

\subsection{Knowledge Guidance}

Prior research has shown that providing LLMs with specialized, domain-specific guidelines enhances accuracy ~\cite{song2025injectingdomainspecificknowledgelarge, yang2023empowerlargelanguagemodel}. In the initial planning step, the Policy agent (\circlednum[blue]{2}) takes as inputs the domain-specific guide and any project-specific policies and constructs a checklist for adaptation guidance. 

{\color{custom-blue}{

\underline{First}, the {\bf domain-specific knowledge} for code adaptation can be derived from the study by Zhang {\em et al.}~\cite{10.1109/ICSE.2019.00046} (Table I). Below is a concrete example of six domain-specific adaptation categories, which are used as the first part of the prompt supplied to the {\bf Policy agent}:

1) {\em Code Hardening:} Enhancing robustness by adding conditionals, handling new exception types, inserting final modifiers, and managing resources (e.g., closing streams).  

2) {\em Resolving Compilation Errors:} Declaring variables, correcting method calls, removing references to undefined methods, etc.

3) {\em Exception Handling:} Modifying \code{try-catch} blocks, updating exception types, and refining \code{catch/finally} logic.

4) {\em Logic Customization:} Adjusting method calls, changing constants or variable types, and refining conditionals for adaptation.  

5) {\em Refactoring:} Improving maintainability (renaming, replacing magic values, restructuring fields).  

6) {\em Other adjustments:} Minor modifications, e.g., altering log statements, reformatting code, or updating annotations and comments. 

\underline{Second}, the {\bf project-specific rules/policies}, if any (e.g., error-handling or naming convention), can be fed into the {\bf Policy agent}. Below is a concrete example of a project-specific list of policies:

1) {\em Exception Naming Convention}: All custom exceptions must follow the suffix \code{*Exception} (e.g., \code{InvalidOrderException}, \code{DependencyResolutionException}) and reflect a clear, domain-specific failure.

2) {\em No Silent Failures}: Exceptions must never be swallowed silently. Every caught exception must be either logged with context or re-thrown with additional information.

3) {\em Methods that may fail should use clear naming conventions}: \code{get*} $\rightarrow$ expected to succeed or throw, \code{tryget*} $\rightarrow$ returns null instead of throwing, and \code{validate*} $\rightarrow$ throws validation-specific exceptions.

4) {\em Resource Cleanup and Fail-Safety}: All methods interacting with external resources (files, network, DB) must ensure proper cleanup (e.g., \code{finally} blocks or \code{try-with-resources}) and maintain system consistency in case of failure.

\underline{Third}, the Policy agent takes as its prompt the two above lists of guidelines and policies, and produces a {\em case‑specific combined checklist}, collectively denoted as \(\mathcal{D}\) (not shown).

As seen in Fig.~\ref{fig:all-prompt} (\(\blacklozenge\)), we then incorporate the case-specific combined checklist \(\mathcal{D}\) at the placeholder \code{<6\_adaptation\_categories> and <policies>}, prompting Planner agent (Section~\ref{sec:plan}) to review the code and methodically determine adaptation categories. If a category is inapplicable, we instruct the model to skip it. The Planner is expected to systematically evaluate each category and determine all potential, necessary changes for the target code \(\mathcal{P}_\mathcal{C}\), rather than focusing on isolated modifications. 

}}

\subsection{Self-Guided Adaptation Plan Generation}
\label{sec:plan}




We build the Adaptation Planner agent (\circlednum[blue]{3}) with LLM and Chain-of-Thought (CoT).
CoT structured reasoning helps LLMs produce more accurate and contextually aligned results when they first develop a structured plan before generating the final solution~\cite{wei2023chainofthoughtpromptingelicitsreasoning, liang2024integratingplanningsingleturnlongform}. 
Our prompt instructs the LLM to build a self-ordered adaptation plan, consisting of executable actions~\(A\) paired with reasoning justifications \(R\). This plan is guided by (1) the previously collected checklist \(\mathcal{D}\) from the Policy agent, (2) the adaptation intent \(\mathcal{I}\) from the Summarizer, (3) the chosen answering code \(\mathcal{P}_\mathcal{C}\), and (4) the target code \(\mathcal{P}_\mathcal{T}\) in a project. The structured reasoning process using CoT can be formulated as:
\begin{align} 
& p(\mathcal{Y} | \mathcal{I}, \mathcal{D},\mathcal{P}_\mathcal{C}, \mathcal{P}_\mathcal{T}) \nonumber & \\ 
& = \prod_{i=1}^{n} p(\mathcal{Y}_i | \mathcal{Y}_1,...,\mathcal{Y}_{i-1}, \mathcal{I}, \mathcal{D},\mathcal{P}_\mathcal{C}, \mathcal{P}_\mathcal{T}) \nonumber & \\
& = \prod_{i=1}^{n} p((\mathcal{A}_i,\mathcal{R}_i) | (\mathcal{A}_1,\mathcal{R}_1),...,(\mathcal{A}_{i-1},\mathcal{R}_{i-1}), \mathcal{I}, \mathcal{D},\mathcal{P}_\mathcal{C}, \mathcal{P}_\mathcal{T}) &
\end{align}
At each reasoning step of CoT, the LLM produces a pair of an action $\mathcal{A}_i$ and a reasoning justification $\mathcal{R}_i$. The decision for a reasoning step depends on the previous reasoning steps and the 4-tuple 
of the intent $\mathcal{I}$, the checklist $\mathcal{D}$, the chosen answering code $\mathcal{P}_\mathcal{C}$, and the target code $\mathcal{P}_\mathcal{T}$. The final output is a plan comprising a \emph{sequence} of pairs of actions $\mathcal{A}_i$ and their corresponding reason~$\mathcal{R}_i$. 


By capturing the rationale alongside the specific action, it ensures the LLM addresses potential changes to be made in its adaptation plan. This plan serves as an explicit blueprint, prompting the model to decompose the adaptation process into distinct, atomic steps \(\mathcal{Y}_i = (\mathcal{A}_i,\mathcal{R}_i)\).
From prior research~\cite{wei2023chainofthoughtpromptingelicitsreasoning, kojima2023largelanguagemodelszeroshot}, such task decomposition reduces inconsistent or incomplete modifications. 


Moreover, by providing \(\mathcal{I}\) (the high-level goal) and \(\mathcal{P}_\mathcal{T}\) (the target environment in which the snippet must integrate), the LLM gains insight into how the new code will function within existing structures. The intent \(\mathcal{I}\) makes sure the generated plan focused on adaptation changes while directly feeding the answering code \(\mathcal{P}_\mathcal{C}\) reveals the context that the plan must address. In combination with domain knowledge in \(\mathcal{D}\), the model can strategize around typical adaptation needs in the target code/project \(\mathcal{P}_\mathcal{T}\) (e.g., handling exceptions or refactoring), thereby maximizing the likelihood of producing coherent, context-appropriate plan \(\mathcal{Y}=(\mathcal{A},\mathcal{R})\).
For example, in the generated plan, shown in highlighted text, Fig.~\ref{fig:all-prompt}(\(\filledstar \)), LLM correctly identifies four categories that need to be adapted in the target code snippet, shown in Fig.~\ref{motiv:example}. It explicitly mentions the potential issue of variable \code{mX}, \code{mY}, \code{TOUCH\_TOLERANCE}, \code{mPath},
\code{circlePath}, \code{mCanvas}, and \code{mPaint}, precisely points out the action it needs to take upon the API call and its usage: {\em "handle both drawing and erasing by properly using \code{.drawPath()}"}.


%% file: technical-section/final-adaptation-phase.tex
\section{Code Adaptation with LLM-Based Patch Generation}
\label{sec:adapt}

Code Adapter agent (\circlednum[blue]{5}) utilizes an LLM to process four key inputs: the generated adaptation plan \(\mathcal{Y} = (\mathcal{A},\mathcal{R}) = \{ (\mathcal{A}_1,\mathcal{R}_1), (\mathcal{A}_2,\mathcal{R}_2),..., (\mathcal{A}_n,\mathcal{R}_n)\}\), the adaptation intent \(\mathcal{I}\), the target code \(\mathcal{P}_\mathcal{T}\), and the context \(\mathcal{C}\). Based on these inputs, the LLM generates a {\em unified diff patch}~\cite{liu2024automatedcodeeditingsearchgeneratemodify} relative to the original code. We opted to prompt the Adapter LLM to {\em produce a diff patch} rather than rewriting the entire file to prevent unintended modifications due to LLM hallucination.  This ensures minimal yet precise modifications to the target code while preserving the original structure~\cite{chen2022codetcodegenerationgenerated}. Moreover, automated program repair (APR) research showed that LLMs are well-suited for generating unified diff patches, enabling seamless integration of changes into source code~\cite{li2024hybridautomatedprogramrepair, liu2024automatedcodeeditingsearchgeneratemodify, xu2025aligningobjectivellmbasedprogram}. The prompt to the Adapter consists of four sections, each serving a specific role in guiding the LLM in adaptation:

1) \emph{Adaptation Plan \(\mathcal{Y}\):}
This provides the LLM with a structured adaptation roadmap generated by the Planner.
The plan explicitly lists modifications and reasons based on adaptation categories.


2) \emph{Code Context $\mathcal{C}$ and Codebase:} Rather than feeding the entire project, we extract the most relevant context, ensuring that the LLM has a precise understanding of where and how the adaptation should be applied. The context $\mathcal{C}$, built by the Context Miner tool (\circlednum[orange]{4}), includes:
(1) \emph{adaptation location (\(\mathcal{C}_\mathcal{L}\))}: The exact points are marked by \code{purpose1} and \code{purpose2} in Fig.~\ref{fig:all-prompt} (Final adaptation prompt) where the changes should be integrated. (2) {\em The file level structure \(\mathcal{C}_\mathcal{G}\) of the codebase} is provided. (3) The list of the API calls \(\mathcal{C}_\mathcal{A}\) that are used in the sibling methods of the adapted method.


3) \emph{Target Code \(\mathcal{P}_\mathcal{T}\):} The specific function or module where the adaptation will occur.

4) \emph{Intent Summarization \(\mathcal{I}\):} This is derived from the first phase.


\vspace{-6pt}
\subsubsection*{Generated Unified Diff Patch}
A patch in the unified diff format consists of multiple hunks $\mathcal{H}$=$\{\mathcal{H}_1, \mathcal{H}_2$ $...\}$, where each hunk represents the modifications localized to an area in consecutive lines.
Formally, the patch generation process can be formulated as:
\begin{align}
& p(\mathcal{H} | \mathcal{I}, \mathcal{Y},\mathcal{C}, \mathcal{P}_\mathcal{T}) \nonumber & \\ 
& = \prod_{i=1}^{n} p(\mathcal{H}_i | \mathcal{H}_1,...,\mathcal{H}_{i-1}, \mathcal{I}, \mathcal{Y}, \mathcal{C}_\mathcal{L}, \mathcal{C}_\mathcal{G}, \mathcal{C}_\mathcal{A},\mathcal{P}_\mathcal{T}) & 
\end{align}
{\em The order of applying the hunks of unified diff patches is {\bf autonomously} decided by Adapter}. 



As shown in Fig.~\ref{fig:all-prompt} (\(\filledstar\)), the adaptation plan ensures that the model correctly differentiates between drawing and erasing when calling \code{drawPath()}. Thus, the LLM explicitly inserts the missing statement:  

\code{+  mCanvas.drawPath(mPath, PaintView.IsPaint ? mPaint : mEraserPaint);}  

This change ensures the function correctly switches between \code{mPaint} (for drawing) and \code{mEraserPaint} (for erasing), addressing a key issue highlighted in the intent \(\mathcal{I}\): {\em "draw and erase parts of an image without losing the background in an Android drawing application."} The LLM also anticipates potential API failures and integrates exception handling for \code{drawPath()} as part of the adaptation plan. 

The resulting adapted code is verified by the external compiler (Verifier \circlednum[orange]{6}). If the code is un-compiled, the error messages are fed to the Adapter agent for correction until a certain limit.



%% file: research-questions/overview.tex
\section{Empirical Evaluation}
\label{sec:eval}

To evaluate {\tool}, we seek to answer to the following research questions:


{\bf RQ1. Correctness Analysis.}
How accurate is the code adapted by {\tool} compared to the actual code adapted by the human developers for online code adaptation? 




{\bf RQ2. Human Adaptation vs {\tool}'s Adaptation.} How does the code adapted by {\tool} align with actual developers' adapted code regarding different adaptation categories?



{\bf RQ3. Ablation Study.} How does each agent of {\tool} contribute to its performance?

{\bf RQ4. Performance with Different Adaptation Complexity.}
How well does {\tool} perform when adapting code with varying complexities regarding numbers of lines and code hunks?



{\bf RQ5. Correctness of Adaptation Plan.} How accurate is the Planner agent in {\tool}?













%% file: research-questions/rq1.tex
\section{Correctness Analysis (RQ1)}
\label{sec:rq1}


\subsection{Experimental Methodology}



\subsubsection{Dataset}
We used Zhang {\em et al.}~\cite{10.1109/ICSE.2019.00046}'s dataset with 629 Java samples of potentially adapted code from SO to GitHub. Each sample includes a SO code snippet and all of its adapted code across multiple GitHub repos. To ensure relevance, we filtered out cases where adaptation support was not needed: we excluded SO posts where discussions revealed that the snippet was self-updated, i.e., the author of the question had independently found and incorporated a solution into the post and later copied the exact original SO code snippet into his codebase. After that, we obtained a~dataset of 542 samples, which gives us 952 pairs of SO code snippets and corresponding Github's adapted code. Among those GitHub repos, we were able to set up and run 201 unique repos with the associated  496 concrete test suites. That gives us a subset  $D_t$ with 259 unique pairs of SO code snippets and GitHub counterparts {\em with test cases}. Let us call the subset {\em without test cases} $D_m$ (693 pairs).





\subsubsection{Procedure and Metrics}
For each data instance, we extracted the question, the answering code snippet that was adapted, and its associated explanatory texts, then fed them into the Summarizer and Planner. We used the corresponding GitHub project from the version preceding the adaptation as the target code and location for the adaptation, and fed them to the Context Miner to extract the sibling methods (as the context) within the same class (as the target code) and all of their API calls. 

 








To assess the correctness of adapted code $P$, we compare it with the post-adaptation code $G$ by developers from the dataset. 
For a sample $s$ in $D_t$ (with test cases), if the adapted code passes all the available test cases for $s$ in the ground truth, we consider it as correct. Otherwise, it is incorrect. {\em Correctness} is the ratio of the instances with all passing test cases over the total instances. 
{\color{custom-blue}{For the samples in $D_m$ {\em without test cases}, we conducted a manual evaluation with two evaluators. Annotators were shown the source Stack Overflow code, the target project context, the model-produced adapted code, and the ground-truth code. For each instance, they assigned a binary label: correct or incorrect. An adaptation was labeled correct only if it satisfied: (1) it implemented the intended functionality of the source snippet in the target setting; (2) it was consistent with the target project context, including relevant variable usages, APIs, and data flows, and (3) it introduced no clear semantic mismatch, such as incorrect API substitution, wrong arguments, or behaviors contradicting the target context. To reduce reference bias, annotators were not asked to judge by similarity to the ground-truth adapted code. Instead, the ground truth was used only as contextual evidence for semantic validity, rather than a direct answer key for textual matching. We resolved disagreements via majority voting with an additional evaluator.
}} 
Moreover, for both $D_t$ and $D_m$, we also compare $P$ and $G$ via three {\em code similarity metrics}: 1) dependency similarity via Program Dependence Graphs (PDGs) (from Joern~\cite{joern-2014}), 2) syntactic similarity via ASTs, and 3) lexical similarity via CodeBLEU and editing similarity.
Our evaluation tool aligns PDG/AST nodes based on their statement types (e.g., assignment, method call), and control/data dependencies (e.g., producer/consumer of variables). An edge in $P$ is considered a match with an edge in $G$ if it has the same statement types at two ends and its edge type (control/data/structure) aligns with that in $G$. {\em F1-score} of the alignment is computed as: {\em precision} is the portion of matched edges in the adapted $P$ over its total edges, and {\em recall} is the ratio of matched edges in the ground-truth $G$ and its~total.

\vspace{-4pt}
\subsubsection{Baselines} GPT-4o is used as the base LLM. With no existing tool, our baselines include:

1) {\em A Chain-of-Thought-(CoT)-only baseline:} This baseline represents a straightforward, LLM-based, minimal-context approach. We provided the SO code snippet to GPT-4o and only a Chain-of-Thought prompt, which instructs the LLM to integrate the SO snippet into the target codebase.

2) {\em Naive prompting baseline:} This single-agent (also using CoT) is similar to the CoT-only baseline except that we also provided to GPT-4o the domain guidance, intent summarization, or context.

3) {\em Two Copy-and-APR baselines:} We simulate a baseline where the SO code is directly copied into the target codebase at the given location, and then an automated program repair (APR) tool is used to fix any errors. The objective is to evaluate if an APR tool can produce a fix as the final adaptation. 
We chose {\em two APR approaches}: Agentless~\cite{agentless2025xia} (a state-of-the-art LLM-based APR)~and GPT-4o (we prompted it to fix any errors). We ran each tool five times and reported the mean values.

\vspace{-2pt}
\subsection{Empirical Results}

\subsubsection{Correctness of Adapted Code}
\label{sec:rq1-correctness}

As seen in Table~\ref{tab:rq1_summary_no_pdg}, in the subset $D_t$ with 259 instances that include test cases, {\tool} achieves {\em higher correctness of 65.2\%}, meaning that 65.2\% of {\tool}-adapted code instances pass all corresponding test cases. 

\begin{wraptable}{r}{3.5in}
\centering
\footnotesize
\setlength\tabcolsep{2.3pt}
\caption{{\color{custom-blue}{Correctness of Adapted Code.}}
Test cases: Pass = apply + build + all tests.
{\em Adaptation Categories on semantically-correct outputs only}.}
\label{tab:rq1_summary_no_pdg}
\vspace{-6pt}
\begin{tabular}{l|c|c||rrrrrr}
\toprule
\multirow{2}{*}{Approach} &
{\textbf{Test cases}} &
{\textbf{Manual-check}} &
\multicolumn{6}{c}{\textbf{Adapt. Category (Counts)}} \\
& Corr. (\%) &
Corr. (\%) &
\textit{CH} & \textit{CE} & \textit{EH} & \textit{LC} & \textit{Ref} & \textit{Misc} \\
\midrule
{\color{custom-blue}{CoT-only}}       & {\color{custom-blue}11.6} & {\color{custom-blue}11.8} & {\color{custom-blue}155} & {\color{custom-blue}62} & {\color{custom-blue}24} & {\color{custom-blue}138} & {\color{custom-blue}63}  & {\color{custom-blue}16}  \\
Naive$^{\star}$             & 29.9 & 35.1 & 128 & 31 & 102 & 438  &  176 &  202 \\
Copy+APR$^{\star}$          & 36.4 & 40.7 &  161 & 46  & 263 &  604 & 287 &  279 \\
Agentless$^{-}$      & NA & 54.0 &
\multirow{2}{*}{198} &
\multirow{2}{*}{78} &
\multirow{2}{*}{377}  &
\multirow{2}{*}{809} &
\multirow{2}{*}{406} &
\multirow{2}{*}{285} \\
Agentless$^{+}$         & 59.7 & NA &
 &  &  &  &  &  \\

\midrule
{\tool} & {\bf 65.2} & {\bf 62.3} & 743 & 183 & 353 & 993 & 432 & 171 \\
\bottomrule
\end{tabular}
\footnotesize

\textbf{Legend:} $^{\star=}$ w/ GPT-4o; AgentLess has two modes: $^{-}$ w/o test cases; $^{+ =}$ w/ test cases.
CH=Code Hardening, CE=Compilation Errors, EH=Exception Handling, LC=Logic Customization, Ref=Refactoring.
\textbf{Note:} Category counts are not mutually exclusive (one instance may trigger multiple categories/types).
\end{wraptable}

For the subset $D_m$ with 693 instances without test cases, 62.3\% of instances are judged correct according to the semantics of the human-adapted ground truth via manual verification. {\color{custom-blue}{We measured inter-rater reliability on the manually checked instances using Cohen’s Kappa. The two annotators achieved an {\em observed agreement of 86.9\%} and a {\em Cohen’s Kappa of 0.72}, indicating substantial agreement.}}
{\color{custom-blue}{Notably, the test cases written by Github developers were aimed to test the code adapted from Stack Overflow and they covered those adapted lines}}. The accuracy on instances without test cases is at a comparable level to that on instances with test cases, suggesting that {\tool}’s performance is stable across the evaluations in two separate subsets.



\subsubsection{Adaptation Categories} 
\label{sec:adaptcat}
Table~\ref{tab:rq1_sem_similarity} indicates that adaptations can still be correct despite being divergent from the human reference. To understand these divergences, we analyzed {\tool}'s changes. Table~\ref{tab:rq1_summary_no_pdg} (right) reports the number of counts per adaptation category produced by our tool. {\color{custom-blue}{We computed them using the categorization tool from Zhang \textit{et al.} ~\cite{zhang-icse19} on adaptation changes}}.

\begin{wraptable}{r}{3.5in}
\centering
\footnotesize
\setlength\tabcolsep{1.25pt}
\caption{{\color{custom-blue}{Code Similarities between Adapted Code and Ground Truth}}}
\label{tab:rq1_sem_similarity}
\vspace{-6pt}
\begin{tabular}{l|c r c c c|c r c c c}
\toprule
\multirow{2}{*}{Approach} &
\multicolumn{5}{c|}{\textbf{Test cases}} &
\multicolumn{5}{c}{\textbf{Manual-check}} \\
& Result & PDG & AST & C.BLEU & Edit
& Result & PDG & AST & C.BLEU & Edit \\
\midrule
\multirow{2}{*}{{\color{custom-blue}{CoT-only}}}
& Pass & {\color{custom-blue}0.45} & {\color{custom-blue}0.74} & {\color{custom-blue}0.61} & {\color{custom-blue}0.54} & Corr   & {\color{custom-blue}0.52} & {\color{custom-blue}0.81}  & {\color{custom-blue}0.65} & {\color{custom-blue}0.54} \\
& Fail & {\color{custom-blue}0.26} & {\color{custom-blue}0.43} & {\color{custom-blue}0.40} & {\color{custom-blue}0.29} & Incorr & {\color{custom-blue}0.32} & {\color{custom-blue}0.45} & {\color{custom-blue}0.46} & {\color{custom-blue}0.35} \\
\midrule
\multirow{2}{*}{Naive$^{\star}$}
& Pass & 0.68 & 0.80 & 0.60 & 0.56 & Corr   & 0.70 & 0.82 & 0.65 & 0.53 \\
& Fail & 0.28 & 0.31 & 0.35 & 0.31 & Incorr & 0.32 & 0.43 & 0.42 & 0.32 \\
\midrule
\multirow{2}{*}{Copy+APR$^{\star}$}
& Pass & 0.71 & 0.80 & 0.57 & 0.53 & Corr   & 0.68 & 0.76 & 0.63 & 0.57 \\
& Fail & 0.27 & 0.39 & 0.26 & 0.25 & Incorr & 0.34 & 0.50 & 0.39 & 0.33 \\
\midrule
\multirow{2}{*}{Agentless$^{-}$}
& Pass & NA & NA & NA & NA & Corr   & 0.73 & 0.78 & 0.61 & 0.58 \\
& Fail & NA & NA & NA & NA & Incorr & 0.40 & 0.51 & 0.46 & 0.40 \\
\midrule
\multirow{2}{*}{Agentless$^{+}$}
& Pass & 0.72 & 0.81 & 0.58 & 0.56 & Corr   & NA & NA & NA & NA \\
& Fail & 0.41 & 0.51 & 0.44 & 0.33 & Incorr & NA & NA & NA & NA \\
\midrule
\multirow{2}{*}{{\tool}}
& Pass & {\bf 0.80} & {\bf 0.86} & {\bf 0.65} & {\bf 0.62} & Corr   & {\bf 0.84} & {\bf 0.87} & {\bf 0.68}  & {\bf 0.62} \\
& Fail & {\bf 0.48} & {\bf 0.64} & {\bf 0.50} & {\bf 0.42} & Incorr & {\bf 0.52} & {\bf 0.65} & {\bf 0.54}  & {\bf 0.44} \\
\bottomrule
\end{tabular}
\footnotesize

\textbf{Legend:} $^{\star=}$ w/ GPT-4o; $^{- =}$ w/o test cases; $^{+ =}$ w/ test cases. PDG and AST: F1-scores for alignments. C.BLEU: CodeBLEU.
\end{wraptable}

\vspace{1pt}
\textbf{\em Code Hardening}. 
{\tool} correctly added defensive programming (e.g., \code{null} checks, API validations) where the ground truth does not. The adapted code passes
all the test cases and even conforms to good coding practice. This improvement
comes from adaptation planning, where explicit guidance was provided to LLM. We confirmed the impact of {\em adaptation planning on~Code Hardening} in our ablation~study. 

\vspace{1pt}
\textbf{\em Exception Handling}. Via do\-main-specific guidelines, the LLM is reminded of critical corner cases, e.g., defensive \code{null} checks, API-specific exceptions, and error handling. 
We also confirmed the {\em impact
of planning to improve Exception Handling} in our ablation study. For example, we found the following correct plan by LLM: "\emph{Step 1: Check for Exception Handling:} {\em \code{ClassNotFoundException} is thrown but not caught within the method. Consider adding handling for this exception... Handle potential null pointer exceptions that may arise from the \code{listFiles()} method if it returns null.} \emph{Step 2: Check for Code Hardening:} {\em Check if directory or \code{packageName} are null and handle appropriately to avoid runtime errors. Ensure the File object represents an actual directory before calling \code{listFiles()}...}"

\vspace{1pt}
{\bf \em Logic Customization}. 
As an example, to adapt a hashing \code{algorithm.getInstance} API call, {\tool} correctly changed the algorithm selection via MessageDigest, shifting from "SHA256" to "SHA1" to match the intent. Moreover, the plan guided the model to adjust the type of \code{sb}, leading it to correctly update from \code{StringBuffer} to \code{StringBuilder}, which is more suitable for single-threaded scenarios. We later showed that {\em intent summary helps improve logic customization}~(Section~\ref{sec:rq3}).

\vspace{2pt}
{\bf \em Resolve Compilation Errors and Refactoring}. 
These tasks seem manageable for modern LLMs. E.g., they perform renaming and extract methods, which might differ from the ground truth.

\vspace{-2pt}
\subsubsection{Similarity of Adapted Code} Table~\ref{tab:rq1_sem_similarity} shows that code similarity metrics for correct adaptations are higher than those for incorrect cases in both $D_t$ and $D_m$. The separation indicates~that static similarity metrics align with both oracles. As a coarse-grained sanity check using the outcome-conditioned means (8 points per oracle), the F1-score for PDG alignment shows the strongest positive association with the "pass" outcome (Pearson $r\approx$0.934 on test outcomes and $r\approx$0.925 on manual outcomes), and is similarly correlated on manual outcomes ($r\approx$0.908). This suggests that {\em code similarity metrics can serve as useful proxy signals for correctness} when test cases are~unavailable.

\vspace{-2pt}
\subsubsection{Comparative Results}
{\color{custom-blue}{
We performed statistical significance testing using McNemar test on paired instance-level outcomes. We formed null hypotheses comparing our tool to each baseline in Table~\ref{tab:rq1_summary_no_pdg}. The resulting p-values are 2.88e-114 for CoT-only variant, 9.07e-21 for Naive*, 2.91e-14 for Copy+APR, and 0.00951 for Agentless. Since all the p-values are below 0.01, we reject the respective null hypotheses, indicating that {\tool} performs better than each baseline.

\underline{First}, comparing {\tool} with all the baselines in Table~\ref{tab:rq1_summary_no_pdg}, because they all use the same underlying LLM (GPT-4o), the observed gains cannot be attributed merely to stronger foundational model capacity. Instead, they indicate the effectiveness of {\em {\tool}'s multi-agent workflow}. \underline{Second}, comparing the result of Naive* with CoT-only's, the gain shows {\em the positive contributions to the performance of adaptation intent summary, domain-specific guidance, and code context}, which confirms our Key Ideas 1--3. Our ablation study (Section~\ref{sec:rq3}) will present the contribution of each individual factor. \underline{Third}, comparing {\tool} with Naive*, we can assess the effectiveness of {\em our multi-agent framework over a single-agent workflow} equipped with the same intent summary, domain-specific guidance, and context. As seen, the multi-agent framework helps improve 35.3\% and 27.2\% more correct adaptation cases over the single agent in $D_t$ and $D_m$, respectively.

Specifically, as seen in Table~\ref{tab:rq1_summary_no_pdg}, {\tool} achieves 63.1\% semantic correctness, i.e., it has {\em 5.5\%-53.6\%} 
and {\em 8.3\%-50.5\%} more correct adaptation cases than the baselines in $D_t$ and $D_m$, respectively.
The advantage is also reflected in static code similarity (Table~\ref{tab:rq1_sem_similarity}): for the correct adaptation code in $D_t$, {\tool} attains higher similarity than every baseline across dependency similarity {\color{custom-blue}{11.1\%–77.8\%}}, syntactic similarity {\color{custom-blue}{6.2\%–16.2\%}}, and lexical similarity {\color{custom-blue}{6.6\%–17.0\%}}. The same observation holds for correct instances in $D_m$, and the same general trend also appears when comparing similarity for incorrect instances. Note that for correctly adapted code (in both $D_t$ and $D_m$), despite not being perfect, it achieves the highest code similarities (dependency, syntactic, and lexical similarities) (Table~\ref{tab:rq1_sem_similarity}). For example, in $D_t$, the code similarities average 80.5\%, 86\%, and 65\%, and in $D_m$ they are 84\%, 87\%, and 68\%, respectively. This suggests that {\tool} produces adaptations that are not only more often correct, but also more structurally and semantically aligned with developers' ground truth. In brief, these performance gaps indicate that {\tool}’s structured adaptation process yields more reliable correctness under both evaluation settings.
}}


\subsubsection{Qualitative Analysis} Table~\ref{tab:rq1_summary_no_pdg} (right) helps explain where these gains come from. The baselines frequently miss key steps such as Refactoring and Resolve Compilation Error, and—without explicit guidance—rarely add Code Hardening or Exception Handling proactively. In contrast, {\tool} performs better on these categories because its guidelines explicitly prompt the LLM to consider code building, robustness checks, and defensive handling, which are often needed.

For Logic Customization, we observe an additional qualitative difference. While the baselines generate a large number of instances in this category, they often compensate in brittle ways: they introduce extra APIs or helper functions to replace missing method invocations rather than using the existing package ecosystem, or they replace meaningful constants with placeholder values. These choices may “patch” compilation or produce superficially plausible code, but they frequently drift away from the intended behavior and from project-consistent API usage. In contrast, {\tool} most consistently performs direct calls to methods already existing in the target project. By leveraging the {\em context summarizer}, {\tool} tends to select calls matching the local style and conventions in the file, leading to better alignment with developers' adapted code.

Another key limitation of the baselines is that they often fail to preserve the {\em adaptation intent} expressed by users. For example, consider a case where a user reports a type error arising from use of the Java class \code{PrintWriter} and the corresponding Stack Overflow answer recommends switching to a \code{FileWriter} wrapped by a \code{BufferedWriter}. Without incorporating this intent, none of the baselines reliably extract or apply the recommendation; they keep using \code{PrintWriter} in the adapted code, which can reproduce the same type error. In contrast, {\tool} is designed to carry intent through the pipeline, enabling it to make the specific substitution that resolves the root cause rather than producing an adaptation that remains behaviorally inconsistent with the user’s goal.




\subsubsection{Unique Adaptation Analysis}
Among all correct adaptation samples (601), we identified {\color{custom-blue}{\em 59 cases uniquely adapted by \tool in which no baseline approach achieved}}. Of these unique ones, 14 occur in the test-suite subset and 45 occur in the manual subset. From these unique instances, we observed that: (1) 40 cases were successfully adapted due to improvements in Logic Customization; (2) 16 cases benefited from stronger Exception Handling; (3) 15 cases were resolved through more effective Code Hardening; (4) 9 cases involved Resolving Compilation Errors; and (5) 21 cases were improved through refined Refactoring (categories can overlap within an instance).


\subsubsection{Limitations} For the {\em incorrect instances} in $D_t$ and $D_m$ by {\tool}, we observe several recurring failures in the adapted code. The most common cause of test failure is that the new project does not build; the top three underlying reasons are (1) API misuse, (2) unresolved dependencies, and (3) signature mismatches. On further inspection, we also find cases where, despite receiving instructions that cover multiple adaptation categories, {\tool} does not fully follow them or introduces incorrect changes for specific categories. We defer to Section~\ref{sec:rq2} the details of these mismatches, especially where {\tool}’s adaptations diverge from real developer changes.

%% file: research-questions/rq2.tex
\section{Tool Adaptation versus Human Adaptation (RQ2)}
\label{sec:rq2}

\begin{figure*}[t]
    \centering
        \includegraphics[width=0.8\textwidth]{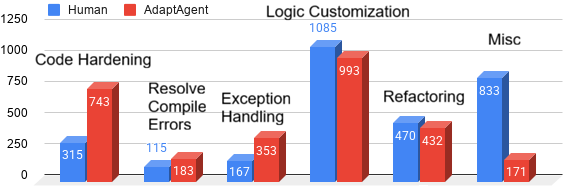} 
    \vspace{-9pt}
    \caption{{\color{custom-blue}{Adaptation Category Distributions}}: Human Adaptation versus Adaptation by \tool (RQ2)}
    \label{fig:human-agent}
\end{figure*}




In this experiment, we study the {\em adaptation effectiveness} of {\tool} by analyzing {\em how close real developers and {\tool} perform various adaptation types}, rather than just the final adaptation. 
To achieve this, we applied Zhang {\em et al.}~\cite{10.1109/ICSE.2019.00046}'s tool to detect different {\color{custom-blue}{adaptation categories}} in both real developer-adapted code and the code produced by {\tool}. 
{\color{custom-blue}{Fig.~\ref{fig:human-agent} displays the counts of different adaptation categories that were performed by GitHub developers and by {\tool}}}. 

\paragraph{Statistical Test}
To assess whether our tool’s adaptation result aligns with that of humans, we compared the distributions of {\color{custom-blue}{adaptation categories}} across the two groups. 
We have two lists of counts, where each number corresponds to the frequency of one of 24 specific adaptation types, with one list derived from human-adapted code and the other from our tool's. To that end, we computed both Spearman’s rank correlation and Pearson’s correlation between the two frequency vectors. If we considered
5 main adaptation categories with respect to their sub-categories (excluding the ``Misc'' main category), Spearman’s test yielded $\rho{=}0.913$ with $p{=}5.1\times10^{-8}$, and Pearson’s test yielded $r{=}0.858$ with $p{=}2.7\times10^{-6}$, both indicating a strong and statistically significant positive association between the two distributions. Thus, {\em the null hypothesis that there is no association between the human and tool adaptation frequencies can be rejected}. This result suggests that {\em our tool is effective in producing adaptation patterns that are {\em strongly aligned}} with those observed in human behavior. If the ``Misc'' main category is included, 
Spearman’s test yielded $\rho{=}0.591$ with $p{=}0.0018$, and Pearson’s test yielded $r{=}0.613$ with $p{=}0.0014$, showing a moderate association. This is expected because the adaptation types
in ``Misc'' (logging, style formatting, comments) are project-specific. 

\paragraph{What are the Differences?}
We analyzed the {\em correctly adapted cases} in both $D_t$ (test cases) and $D_m$ (manual-check) in which {\em the numbers of adaptation types by the tool and humans are different}. We observed that our tool produces useful code including exception handling, defensive checks, etc. (similar to the results in Sections~\ref{sec:rq1-correctness} and~\ref{sec:adaptcat}), which still passes all the test cases. These extra edits typically close resources or wrap I/O in \code{try\{...\}catch\{...\}} blocks, which improve robustness, but also introduce false-positive edges, explaining why code similarity is not perfect.
Moreover, as seen Fig.~\ref{fig:human-agent}, our tool suggests more of the following adaptation types than actual humans' adaptation, many of which are useful: 1) handle a new exception type (98 vs 60), 2) clean-up unmanaged resources (4 vs 3), 3) insert \code{try-catch} (100 vs 52), 4) change \code{catch} blocks (98 vs 47) and 5) add conditions (364 vs~167). The analysis on additional changes by {\tool} is presented later.

In comparison, Agentless performs 809 logic customizations (184 fewer than \tool and 276 fewer than developers) while inflating \emph{Miscellaneous} changes (e.g., comments, logging, formatting) to 285, much higher than 171 in \tool. At the same time, the shortfall in logic customizations means Agentless fails to recover many of the true behavioral dependencies.



\paragraph{Additional Adaptation Changes from {\tool}}

This section presents a targeted analysis of whether {\tool}'s additional adaptation changes in Code Hardening (CH) and Exception Handling (EH) are useful in {\em preventing run-time errors/crashes}. For Code Hardening, humans applied 315 changes, whereas \tool produced 743, i.e., 428 extra changes. For Exception Handling, humans~applied 167 changes, whereas \tool produced 353, more than the human-written adaptations. To assess to what degree these additional changes are necessary, we conducted an run-time experiment as follows. We first identified the set of Code Hardening changes introduced by {\tool} but {\em not} present in the respective human-adapted snippets. We then retained only those snippets whose target projects are compilable and have executable test cases, yielding 188 unique extra CH instances. We performed the same procedure for Exception Handling, obtaining 106 unique extra EH instances. We denote the resulting sets by $D_{CH}$ and $D_{EH}$,~respectively. 

 

For each code snippet $c_e \in D_{CH}$, we identified its corresponding human-adapted code $c_o$ in our dataset, where the Code Hardening change by {\tool} is absent. We then prompted GPT-4o with each pair of snippets ($c_o$, $c_e$) to generate an input that would trigger an error in $c_o$ but not in $c_e$ thanks to the extra Code Hardening, thereby testing whether the added hardening logic prevents a failure that could otherwise occur. Similarly, for each code snippet in $D_{EH}$, we prompted GPT-4o to generate an input that would trigger an exception in the human version $c_o$ but not in the tool-adapted version $c_e$, thereby evaluating the necessity and correctness of the added Exception Handling logic. We repeated each prompting process at most 3 times and reported the number of cases in which the enhanced version $c_e$, augmented with the additional Code Hardening or Exception Handling logic, successfully avoided the runtime errors or exceptions triggered in $c_o$.

Table~\ref{tab:additional-ch-eh} shows that a substantial portion of the additional adaptation logic by \tool is indeed useful. For Code Hardening (CH), in {\em 40.3\%} of the 188 extra CH instances, the enhanced version $c_e$ successfully avoided a runtime failure that occurred in the human version $c_o$, showing that the added hardening logic was needed for the execution under realistic inputs. For Exception Handling (EH), this effect was observed in {\em 44.8\%} of the 106 extra EH instances, where the added exception-handling logic in $c_e$ successfully prevented an exception triggered in $c_o$. These results suggest that the extra CH and EH changes by {\tool} often provide meaningful robustness improvements. Overall, they show {\tool}'s usefulness in {\em generating necessary Code Hardening and Exception Handling logic beyond what might actually be missed by~humans}.

\input{figures/CH_EH}

%% file: figures/CH_EH.tex
  \begin{table}[t]
  \centering
  \small
  \caption{{\color{custom-blue}Usefulness of additional Code Hardening and Exception Handling code introduced by {\tool}}.}
  \label{tab:additional-ch-eh}
  \vspace{-9pt}
  \begin{tabular}{lcc}
  \toprule
  Adaptation Category & Additional CH or EH instances & Errors Triggered \\
  \midrule
  Code Hardening & 188 & 40.3\% \\
  Exception Handling & 106 & 44.8\% \\
  \bottomrule
  \end{tabular}
  \end{table}

%% file: research-questions/rq3.tex
\section{Ablation Study (RQ3)}
\label{sec:rq3}

{\tool} consists of four critical agents: the adaptation plan \(\mathcal{Y}\) via {\bf Planner}, adaptation intent \(\mathcal{I}\) via {\bf Summarizer}, the code context $\mathcal{C}$ via {\bf Cxt Miner}, and {\bf Adapter} agent. Because Code Adapter cannot be removed, to systematically evaluate the importance of each agent, we removed each of other agents separately, creating three variants: 1) without Planner, 2) without Summarizer, and 3) without context.
Table \ref{tab:ablation} and Fig.~\ref{fig:type distribution RQ3} show how each agent uniquely impacts adaptation performance. 

\subsection{Impact of Agents}


We performed statistical significance test on Table~\ref{tab:ablation} using the McNemar test on paired instance-level outcomes. The p-values are 6.84e-05 for “w/o Planner”, 1.79e-07 for “w/o Summarizer”, and 6.12e-04 for “w/o Context Miner”. Since all p-values are below 0.01, we reject the corresponding null hypotheses, concluding that the ablation variants performed worse than {\tool}.

\input{research-questions/rq3-table}


\subsubsection{Adaptation Plan via \underline{Planner} agent}

The removal of the {\em Planner agent from our pipeline significantly decreases correctness in the Code Hardening and Exception handling} categories, making them similar to naive prompting's results. The analysis on adaptation types in Fig.~\ref{fig:type distribution RQ3} reinforces this observation: without adaptation planning, the adaptation missed more on crucial adaptation types like ``add a conditional,'' ``handle a new exception type,'' and ``insert/delete \code{try-catch} blocks''. Comparing Fig.~\ref{fig:human-agent} and Fig.~\ref{fig:type distribution RQ3}, these adaptation types observe a sharp decline of 33.5\%–75.4\%. This confirms our idea that structured planning guides LLMs to perform critical adaptation steps on Code Hardening and Exception Handling.



\subsubsection{Adaptation Intent via \underline{Summarizer} agent}

\begin{figure}[t]
    \centering
    \includegraphics[width=3.6in]{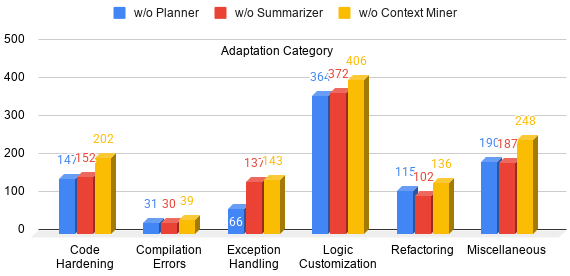} 
    \vspace{-9pt}
    \caption{{\color{custom-blue}{Adaptation Type Distribution in Ablation Study (RQ3)}}}
    \label{fig:type distribution RQ3}
\end{figure}

Without incorporating adaptation intent $\mathcal{I}$ created by the Summarizer agent, performance in {\em Logic Customization} notably drops by 30.8\%--70.8\% more compared to the two other variants. Fig.~\ref{fig:type distribution RQ3} shows a decrease in the adaptation type ``change method~call'', indicating that {\em the model relies heavily on the high-level, abstract information provided by the intent to modify API calls or conditional logic accurately}. 



\subsubsection{Structural Information of Codebase and Context via \underline{Context Miner} agent}

Excluding structural context (sibling methods of the target code), considerably affects adaptations, evident in adaptation types such as ``change method call,'' ``specify target of method call,'' and ``change access modifiers.'' Comparing Fig.~\ref{fig:human-agent} and Fig.~\ref{fig:type distribution RQ3} demonstrates a clear drop in these adaptation types, nearly comparable to the absence of intent. The correctness does not drastically decline. However, without structural context via Context Miner, the LLM lacks essential guidance on program semantics.






\subsubsection{Combination}

Our results also reveal the scenarios where adaptation intent via {\bf Summarizer} agent and code context via {\bf Context Miner} agent have {\em compound impacts}. For example, when both intent and code context are provided (i.e., w/o Planner in Fig.~\ref{fig:type distribution RQ3}), the category ``change method call'' achieves the highest frequency among the three variants, indicating that these two agents significantly aid the LLM in correctly identifying necessary APIs. Similarly, in adaptations involving ``type resolution,'' both intent and context independently yield comparable improvements, meaning better understanding of the user package and type resolution due to the awareness of the target~code. 



Our results show that {\em the adaptation plan with {\bf Planner} serves as a structured template guiding LLMs' reasoning}, while {\em adaptation intent with {\bf Summarizer} and code context with {\bf Ctx Miner}~provide complementary high-level and low-level details}, helping the model make correct adaptations.




\subsection{Comparison on Impacts of Agents}


\subsubsection{Planner vs Summarizer}

For tasks like ``add thrown exceptions'' ``change method calls,'' and ``declare a type,'' Planner seems more critical than Summarizer. This suggests that {\em while understanding what the user wants (intent) is important, knowing how to achieve it (planning) is even more crucial for these specific adaptations, as Summarizer is most critical for {\em logic changes} (e.g., method-call changes)}.



\subsubsection{Summarizer vs Context Miner}

The Summarizer can drive Logic Customization, indicating that the LLM benefits more from capturing the user’s adaptation goal than from additional target-code context, improving strongly in correctness via test cases. In contrast, Ctx Miner can support the  integration for tasks such as renaming, inserting a \code{final} modifier, updating exception types.



%% file: research-questions/rq3-table.tex
\begin{wraptable}{r}{3.5in}
\centering
\footnotesize
\setlength\tabcolsep{2.3pt}
\caption{Ablation Study on Correctness (RQ3).
Test cases: Corr. = apply +build+all tests.
{\em Adapt. Categories on semantically-correct outputs only}.}
\label{tab:ablation}
\vspace{-6pt}
\begin{tabular}{l|c|c||rrrrrr}
\toprule
\multirow{2}{*}{Variant} &
{\textbf{Test cases}} &
{\textbf{Manual-check}} &
\multicolumn{6}{c}{\textbf{Adapt. Category (Counts)}} \\
& Corr. (\%) &
Corr. (\%) &
\textit{CH} & \textit{CE} & \textit{EH} & \textit{LC} & \textit{Ref} & \textit{Misc} \\
\midrule
w/o Planner      & 52.4 & 51.0 & 173 &  38 & 257 & 728 & 307 & 229 \\
w/o Summarizer   & 46.7 & 47.6 & 558 & 139 & 276 & 511 & 322 & 151 \\
w/o Ctx Miner    & 49.8 & 56.8 & 627 & 158 & 368 & 706 & 284 & 193 \\
\bottomrule
\end{tabular}
\footnotesize

\textbf{Legend:} Variants remove one component from {\tool}:
Planner (adaptation plan), Summarizer (intent), or Ctx Miner (semantic context).
CH=Code Hardening, CE=Compilation Errors, EH=Exception Handling, LC=Logic Customization, Ref=Refactoring.
\textbf{Note:} Category counts are aggregated type-level edits; counts are not mutually exclusive.
\end{wraptable}

%% file: research-questions/rq4.tex


\section{Stratified Performance Analysis on Adaptation Complexity (RQ4)}
\label{sec:rq4}

\input{research-questions/sensitivity-results}


In this experiment, we assess how well our tool performs on the adaption changes involving multiple lines and hunks of code. 
We stratified the result in RQ1 into different groups with respect to numbers of lines and numbers of hunks in the diff patches returned from {\tool}.


\paragraph{Performance by Adaptation Complexity in terms of LOCs}

The smallest adaptation change involved 4 LOCs, while the largest spanned 220 lines, with an average of 42 lines per instance. As seen in Table~\ref{tab:rq4linenumber}(A), our tool achieves 79.9\% correctness for minimal adaptation changes (<10 lines) and maintains high scores for medium changes (71.7\% for 11-20 lines, 68.0\% for 21-30 lines, 65.1\% for 31-40 lines, and 61.8\% for 41-50 lines). However, the correctness drops much to 58.3\% for the adaptations with 51-60 lines and drops to 46.7\% for adaptation changes exceeding 71~lines. Interestingly, the 61–70 line range achieves a higher correctness despite being longer because its 5 instances involve comparably simpler edits.


To further understand this performance drop, we conducted a category-wise analysis. For changes under 50 lines, \tool typically handles only 1.6 adaptation categories with on average 3.6 adaptation types on average. However, for changes exceeding 50 lines, the number of adaptation categories increases to 2.9, involving an average of 7.9 adaptation types per instance. This makes the adaptation significantly more complex. Notably, for large code, \tool struggles the most with Logic Customization, particularly when adapting API calls into the target code.


\vspace{-5pt}
\paragraph{Performance by Adaptation Complexity in terms of Hunks}

Table \ref{tab:rq4pythoncode}(B) presents {\tool}'s performance across different patch complexities, measured by the number of hunks of code changes for adaptation. For single-hunk adaptation ($N$=1), it achieves an correctness of 73.3\%. Each hunk modifies between 1--103 lines of code, with an average modification of 18 lines. Thus, it performs well even with a large number of consecutive lines for a single-hunk adaptation.

When the adaptation involves two hunks ($N$=2), the performance drops from 73.3\% to 64.6\%. This is because most two-hunk patches involve modifying \code{import} statements and implementing the function body. However, with three hunks ($N$=3), the performance drops significantly to 51.9\%, and further declines to 40.1\% at $N$=4. For patches with more than four hunks ($N$>=5), it drops to 33.3\%. This decline occurs because, in more complex adaptations, \tool starts modifying unrelated sections of the method. This decline correlates with the structural complexity of the adaptation. For instance, the average code length at $N$=1 is 43 lines, while at $N\ge$ 5 it increases to 61 lines, with a 41.9\% increase. However, during the same progression, the correctness decreases from 73.3\% to 33.3\%, a relative 120\% decrease.

This suggests that when edits are localized (fewer hunks), the model is better at preserving the structural alignment implied by the intent. However, as adaptations spread across more hunks, especially into unrelated blocks in the method, the model struggles to maintain consistent adaptation.

%% file: research-questions/sensitivity-results.tex
\begin{table}[t]
\centering
 \caption{Correctness by Line Number and by Number of Code Hunks in Adapted Code (RQ4)}
  \label{tab:rq4linenumber}
\begin{minipage}[t]{0.48\textwidth}
  \centering
  \small
  \vspace{-9pt}
  \tabcolsep 3pt
    \begin{tabular}{c|c|c|c|c|c|c|c}
    \hline
    \multicolumn{8}{c}{(A) \textbf{Performance by Line Numbers}} \\ \cline{1-8}
    \textbf{<10}   & \textbf{11-20}  & \textbf{21-30}  & \textbf{31-40}  & \textbf{41-50} & \textbf{51-60}  & \textbf{61-70} & \textbf{>71} \\ \hline
    {79.9} & {71.7} & {68.0}  & {65.1}   & {61.8} & {58.3} & {53.4} & {46.7}    \\ \hline
  \end{tabular}
\end{minipage}
\hfill
\begin{minipage}[t]{0.48\textwidth}
  \centering
  \small
  \label{tab:rq4pythoncode}
  \tabcolsep 3pt
  \vspace{-9pt}
    \begin{tabular}{c|c|c|c|c}
    \hline
    \multicolumn{5}{c}{(B) \textbf{Performance by Number of Hunks}} \\ \cline{1-5}
    \textbf{$N$=1}   & \textbf{$N$=2}  & \textbf{$N$=3}  & \textbf{$N$=4}  & \textbf{$N$=5+}   \\ \hline
    {73.3} & {64.6} & {51.9}  & {40.1}   & {33.3}    \\ \hline
  \end{tabular}%
\end{minipage}
\end{table}

%% file: research-questions/rq5.tex
\section{Correctness of Adaptation Plan Generated by the Planner Agent (RQ5)}
\label{sec:rq5}



\begin{wraptable}{r}{3.3in}
\small
\caption{{\color{custom-blue}{Quantitative Analysis on Micro Reasoning of Planner and Adapter. CH: Code Hardening, CE: Compiling Errors, EH: Exception Handling, LC: Logic Customization, REF: Refactoring.}}}
\label{tab:plan}
\vspace{-6pt}
\begin{tabular}{cc|l|l|l|l|l|l}
\toprule
\multicolumn{2}{l|}{\begin{tabular}[c]{@{}l@{}}Adaptation Type$\rightarrow$ \\ Agent$\downarrow$\end{tabular}} & CH & CE & EH & LC & REF & MISC \\ \midrule
\multicolumn{1}{l}{\textbf{Planner}}         & correct                        & {\color{custom-blue}179} & {\color{custom-blue}199} & {\color{custom-blue}147} & {\color{custom-blue}246} & {\color{custom-blue}186}  & {\color{custom-blue}165}   \\ \hline
\multirow{3}{*}{\textbf{Adapter}}              & correct                      & {\color{custom-blue}165} & {\color{custom-blue}190} & {\color{custom-blue}134} & {\color{custom-blue}187} &  {\color{custom-blue}160} &  {\color{custom-blue}140}  \\ \cline{2-8} 
                                            & missing                         & {\color{custom-blue}14}  & {\color{custom-blue}9}  & {\color{custom-blue}13}  & {\color{custom-blue}59}  &  {\color{custom-blue}26}  &  {\color{custom-blue}25}   \\ \cline{2-8} 
                                                & extra                       & {\color{custom-blue}77}  & {\color{custom-blue}23}  & {\color{custom-blue}42}  & {\color{custom-blue}62}  &  {\color{custom-blue}31}  & {\color{custom-blue}22}  \\
\bottomrule
\end{tabular}
\end{wraptable}

In this experiment, we evaluate the correctness of the generated adaptation plans by the Planner agent regarding the six adaptation categories. {\color{custom-blue}{To achieve the confidence level of 95\% with the margin of error of 5\%, and the population proportion of 50\%, we randomly selected {\em 274} adaptation instances among 952 instances in RQ1 for manual checking by an independent evaluator.
}}
The model's reasoning in texts (see Fig.~\ref{fig:reasoning} for an example) was assessed to validate if the reasoning correctly constitutes any adaptation category in Zhang {\em et al.}~\cite{10.1109/ICSE.2019.00046}. Specifically, for the planning steps from the Planner, (s)he counted the number of incorrect instances in planning if the plan misclassified a needed change, omitted a needed step, added an unnecessary step, or sequenced steps incorrectly relative to the ground~truth. 

As seen in Table~\ref{tab:plan}, the planning steps for Logic Customization are the most correct ({\color{custom-blue}{246/274}}), while Exception Handling is more challenging for the Planner ({\color{custom-blue}{147/274}}). In Logic Customization, instead of correctly removing an undefined variable, it attempts to initialize and integrate it into the existing method, potentially altering the original functionality. For Code Hardening, Compiling Errors, and Refactoring, the Planner performs reasonably well ({\color{custom-blue}{179-199 over 274}}).

Additionally, each portion of the generated adaptation of each instance (in generated source code) is mapped with each planning step (in texts) in each adaptation category to examine {\em the cases where the Planner provides correct reasoning, yet the Adapter does not produce the correct adapted code}. 
{\color{custom-blue}{The Adapter adapts Compilation Errors, Code Hardening, and Exception Handling with relatively high fidelity, while Logic Customization has the largest missing gap.}}


\input{research-questions/plan-example-2}

In Table~\ref{tab:plan}, among {\color{custom-blue}{179}} correctly planned instances for Code Hardening, {\color{custom-blue}{165}} of them have the correctly generated adaptation code. {\color{custom-blue}{14}} instances have missed Code Hardening, while {\color{custom-blue}{77}} of them have extra ones. For Exception Handling, among {\color{custom-blue}{147}} correctly planned instances, {\color{custom-blue}{134}} are correctly adapted, while {\color{custom-blue}{13}} are missed and {\color{custom-blue}{42}} extra Exception Handling are introduced.

We further investigated the error patterns from the Adapter. A common error in Logic Customization is that a variable should have been declared at the file level; however, the Adapter added it as a final modifier, a local variable,  or an input parameter. The other errors include ``not updating String value'', ``missing repo-level APIs'', and ``incorrect changing types''. For Exception Handling, the common mistake is an exception type mismatch with the ground truth.

%% file: research-questions/plan-example-2.tex
\begin{wrapfigure}{r}{3.2in}
	\centering
	\lstset{
		numbers=left,
		numberstyle= \tiny,
		keywordstyle= \color{blue!70},
		commentstyle= \color{red!50!green!50!blue!50},
		rulesepcolor= \color{red!20!green!20!blue!20} ,
		xleftmargin=1.5em,xrightmargin=0em, aboveskip=1em,
		framexleftmargin=1.5em,
        framexrightmargin=-1em,
        numbersep= 5pt,
		language=Python,
    basicstyle=\scriptsize\ttfamily,
    numberstyle=\scriptsize\ttfamily,
    emphstyle=\bfseries,
                moredelim=**[is][\color{red}]{@}{@},
		escapeinside= {(*@}{@*)}
	}
\vspace{-18pt}
\begin{lstlisting}[]
1. **Code Hardening**:
- Add validation to check if the `encrypted` byte array is null or empty.
Validate that `keyIn` is not null before attempting to use.
- Ensure the `password` array is not null before calling `findSecretKey` .
- Handle potential resource leaks by making sure `InputStream` and `ByteArrayOutputStream` are closed in a finally block or using try-with-resources. ...
3. **Exception Handling**:
- Wrap the method body in a try-catch block to handle specific exceptions and provide meaningful error messages.
- Consider handling `PGPDataValidationException` separately if it can occur in this context.
- Ensure that `PGPException` and `NoSuchProviderException` are appropriately handled if they are not directly thrown by this method but during library usage....
5. **Refactoring**:
- Ensure that `sKey` and `pbe` are properly initialized with values to prevent null references before usage.
- Rename variables to more descriptive names, for instance, rename `in` to `decodedStream`, `o` to `decryptObj`.
- Break down the method into smaller private helper methods to increase readability, such as `getPrivateKey`, `getPgpObjectFactory`, etc....
\end{lstlisting}
\vspace{-18pt}
\caption{A Natural-language Adaptation Plan by Planner Agent}
\label{fig:reasoning}
\end{wrapfigure}

%% file: limitations.tex
\section{Threats to Validity}


\paragraph{Internal Validity}
A potential threat concerns data leakage. We showed the results of both GPT-4o with naive prompting and {\tool}, which increases the total number of correct adaptations in all types. This shows that the improvement stems from our workflow rather than GPT-4o’s pre-trained knowledge alone. GPT-4o's underperformance suggests a low risk of data leakage. In our dataset, not all the code instances have test cases, we had to manually check the subset $D_m$.



\paragraph{External Validity}
While our Java dataset may not be representative, it was comprehensively collected by Zhang {\em et al.}~\cite{10.1109/ICSE.2019.00046} from real-world SO posts and humans' adaptations. Similarly, we evaluated only on Java and GPT-4o, and results may vary with other models on other languages. For example, our implementation did not support Python thus we did not use AdaptEval dataset~\cite{zhangase25adapteval}. 

\paragraph{Construct Validity}
Another risk lies in the design of prompts, whose variations in wording and level of detail can affect the results. For mitigation, we standardized prompts across models. To handle non-determinism, we executed the LLMs five times and reported the mean values.


We rely on the posts' content, making it dependent on their quality. It may under-perform when SO posts lack sufficient context or when adaptation involves complex domain-specific reasoning. For instance, if the code switches from one API to another,
without mentioning it in the text, the LLM cannot infer the intent. The adaptions requiring multiple hunks are also challenging. 



%% file: related.tex
\section{Related Work}
\label{sec:related}


\paragraph{Studies on quality of SO code examples}

Verdi {\em et al.}~\cite{verdi-tse22} analyzed 1,325 SO answers and found 99 vulnerable snippets that propagated to 2,589 GitHub repositories. \cite{hong21dicos} detected 14,719 insecure code snippets from 1,958,283 SO posts, confirming that 151 out of 2,000 popular C/C++ projects contained at least one insecure snippet. Similarly, 15.4\% of 1.3 million Android apps were found to include security-related SO code~\cite{fisher17stack}. Ragkhitwetsagul {\em et al.}~\cite{chaiyong2021toxic} also reported that S/O code imported into GitHub projects often contains defects, vulnerabilities, performance issues.



\vspace{-2pt}
\paragraph{Studies on Online Code Adaptation}

Zhang {\em et al.}~\cite{10.1109/ICSE.2019.00046} manually examine SO snippets and their GitHub counterparts, proposing a taxonomy of 24 adaptation types across 6 categories. Zhang {\em et al.}~\cite{zhang2024how} analyze 300 code reuse cases with 1,384 adaptations and find that adaptation patterns are repetitive, which benefits automatic adaptation. Li {\em et al.}~\cite{li2022debugging} report that developers' ability to identify helpful posts for debugging varies based on expertise levels.


\vspace{-2pt}
\paragraph{Tool supports for SO code examples}
Cai {\em et al.}~\cite{icse24-exception} develop a BERT-based tool to automatically wrap SO code snippets with \code{try-catch} blocks and handle exceptions. Huang {\em et al.}~\cite{huang2023prompt, huang2023fqn} fine-tune a code masked language model (MLM) for type inference in statically-typed partial code using a “pre-train, prompt, and predict” paradigm. Venigalla {\em et al.}~\cite{venigalla2021stackemo} introduce StackEmo, a Chrome plugin that adds sentiment-based emojis to SO comments. Horvath {\em et al.}~\cite{horvath2024meta} create Meta-Manager, which tracks and organizes code changes, including AI-generated and copy-pasted code. Mahajan {\em et al.}~\cite{mahajan2020recommending,mahajan2022providing}'s MAESTRO suggests the most relevant SO post for a given Java RE. TechSumBot~\cite{yang2023answer} aids in usefulness ranking, centrality estimation, and redundancy removal for technical Q\&A.
NLP2TestableCode~\cite{reid2020optimising} search for code in SO and modify snippets to improve fit and correct errors.

{\color{custom-blue}For code adaptation, one could copy a code snippet into a codebase and leverage an automated program repair (APR) tool for correction. We compared {\tool} with such an approach in our experiments. According to the survey by Huang {\em et al.}~\cite{huang2024evolving}, learning-based APR is better than traditional non-AI APR on some dimensions, especially lower patch overfitting and better generality. For example, ARJA~\cite{yuan2020arja} uses the partial fixes in the same codebase. Moreover, ARJA requires the test cases, which are available for about 25\% of our dataset.}



%% file: conclusion.tex
\vspace{-2pt}
\section{Conclusion}



We present {\tool}, a knowledge-guided, intent-aware, and context-driven  {\bf agentic framework} that leverages LLMs to automate code adaptation from online forums to a target codebase. Our findings show that simply prompting LLMs to adapt a given code snippet yields suboptimal results, often missing critical modifications. By incorporating adaptation {\bf guidelines}, capturing adaptation {\bf intent} via textual summarization, and providing the semantic context, we enable LLMs to generate a structured and accurate {\bf adaptation plan}, leading to better adaptation.

\vspace{-3pt}
\paragraph{Significance and Impact}

\underline{First}, our {\bf agentic framework} to integrating domain-specific guidance and semantic context can be extended to broader {\bf software maintenance tasks that involve code changes}, such as automatic program repair, automated code refactoring, and API or framework migration, etc. \underline{Second}, its structured LLM-based planning for code changes can enhance AI-assisted IDEs by improving reasoning about multi-step modifications, suggesting code changes. \underline{Third,} our principles can be applied to improve program repair by guiding LLMs to understand the root causes of defects and propose context-aware bug-fixing change plans and actual fixes. Finally, the cost of our approach is also reasonable. Our experiment costs \$47 (in terms of API calls, latency, and token consumption), i.e., \$0.05 per instance. The cost of naive prompting is \$14.7, i.e., \$0.018 per instance.